\documentclass[preprint,12pt]{elsarticle}

\usepackage{graphicx}
\usepackage{amssymb}

\journal{Journal of Quantitative Spectroscopy and Radiative Transfer}

\begin{document}

\begin{frontmatter}


\title{Mapping the brightness and color of urban to rural skyglow with all-sky photometry}



\author[label1,label2]{Andreas Jechow}
\author[label2,label1]{Christopher C. M. Kyba}
\author[label1,label3]{Franz H{\"o}lker}

\address[label1]{Ecohydrology, Leibniz Institute for Freshwater Ecology and Inland Fisheries, M{\"u}ggelseedamm 310, 12587 Berlin, Germany}
\address[label2]{Remote Sensing and Geoinformatics, GFZ German Research Centre for Geosciences, Telegrafenberg, 14473 Potsdam, Germany}
\address[label3]{Institute of Biology, Freie Universität Berlin, Berlin, Germany}

\begin{abstract}
Artificial skyglow is a form of light pollution with wide ranging implications on the environment. The extent, intensity and color of skyglow depends on the artificial light sources and weather conditions. Skyglow can be best determined with ground based instruments. We mapped the skyglow of Berlin, Germany, for clear sky and overcast sky conditions inside and outside of the city limits. We conducted observations using a transect from the city center of Berlin towards a rural place more than 58 km south of Berlin using all-sky photometry with a calibrated commercial digital camera and a fisheye lens. From the multispectral imaging data, we processed luminance and correlated color temperature maps. We extracted the night sky brightness and correlated color temperature at zenith, as well as horizontal and scalar illuminance simultaneously. We calculated cloud amplification factors at each site and investigated the changes of brightness and color with distance, particularly showing differences inside and outside of the city limits. We found high values for illuminance above full moon light levels and amplification factors as high as 25 in the city center and a gradient towards the city limit and outside of the city limit. We further observed that clouds decrease the correlated color temperature in almost all cases. We discuss advantages and weaknesses of our method, compare the results with modeled night sky brightness data and provide recommendations for future work.
\end{abstract}


\end{frontmatter}


\section{Introduction}
Light pollution (LP) is caused by excessive use of artificial light at night (ALAN) \cite{Riegel1973}. Over the last decades, LP has become a major concern as environmental stressor for flora and fauna \cite{Longcore2004,book:rich_longcore,Gaston:2015_ptb,Hoelker:2010_b} with potential effects on human health \cite{Stevens:2015}. On a global average, ALAN sensed from space is growing exponentially in area and intensity by more than 2 $\%$ per year, while some countries show higher growth rates of more than 10 $\%$ per year \cite{kyba2017VIIRS}. Furthermore, ALAN is undergoing a color change from gas discharge lamps to modern solid state technology \cite{kyba2017VIIRS}.

Artificial skyglow is a form of LP that is caused when ALAN is directly or indirectly radiated towards the sky and then scattered within the atmosphere. This results in an increased night sky brightness (NSB) \cite{Aube:2015}. The intensity, color and spatial extent of skyglow change with weather conditions, most dramatically with clouds \cite{Kyba:2015_isqm,Kyba:2011_sqm,Ribas2016clouds,jechow2017balaguer} and snow \cite{jechow2019snowglow}.

In contrast to the localized LP from direct light emission, skyglow can affect large spatial scales because photons can travel long distances before a scattering event occurs. A global skyglow model predicts that 80$\%$ of the US and 60$\%$ of the EU population live under skies so much light polluted, that they cannot see the Milky Way \cite{falchi2016WA}. There is evidence, that skyglow can have an impact on whole ecosystems and biodiversity \cite{kyba2013artificially}. Furthermore, it was shown that typical (clear sky) skyglow level light intensities can have impact on the hormone system of several vertebrates (e.g. fish and rodents) \cite{grubisic2019light}. Experimental work on the biological impacts of skyglow is difficult, but a study on zooplankton in an urban lake showed that skyglow can have an impact on a freshwater ecosystems\cite{Moore2000}. These environmental impacts potentially also increase when skyglow is amplified by clouds \cite{Kyba:2011_sqm,jechow2017balaguer}. Furthermore, the color information is becoming more and more important for ecological studies and recent work suggests that the spectral information is crucial \cite{longcore2018rapid}. Thus, it is highly desired to know the status quo of skyglow in terms of NSB, illuminance and color for clear and cloudy conditions, which is best determined by ground based measurements.

The objectives of this work were i) to examine the night sky brightness, the illuminance and the color of the night sky as a function of distance from the center of a large city, and ii) to investigate the influence of clouds on the changes of those parameters. To achieve this, we map the brightness and color of the skyglow of the city of Berlin for clear sky and overcast sky conditions inside and outside of the city limits. We performed transects from the city center of Berlin towards a rural place more than 58 km south of Berlin using all-sky photometry with a calibrated commercial digital single lens reflex (DSLR) camera and a 190$^{\circ}$ fisheye lens. This method has several advantages compared to single-channel devices, that particularly lack color information. DSLR photometry provides full spatial information over the imaging hemisphere, multispectral color information from 3 bands (RGB), it is easy to operate and very well suited for field work \cite{Kollath:2010,kollath2017night,Jechow2016,jechow2017measuring,jechow2017balaguer,jechow2019observing,jechow2019snowglow}. 

From the imaging data, we derived luminance maps and extracted zenith brightness, illuminance and correlated color temperature (CCT) information simultaneously. We calculated cloud amplification factors at each site and investigated the changes in brightness and color with distance, particularly showing differences inside and outside of the city limits. We also investigated the ratio between zenith luminance and illuminance for each measurement, which is important to infer illuminance from zenith brightness measurements. We compare our results with a global skyglow model, discuss advantages and weaknesses of our method and provide recommendations for future work.

\section{Methods}
\subsection{Study site}
To map the skyglow of Berlin in one direction, a transect with twelve stops from the city center towards a rural area was performed. Five stops were within and seven outside of the city limits. The dataset was obtained during the clear night from 28.-29.03.2017 between ca. 1:30 a.m. and 4:45 a.m. local time on 29.03.2017 and during the overcast night from 29.-30.03.2017 at about the same time. For the overcast night, an additional stop was made at Berlin Alexanderplatz (stop 0). The transect was designed along the line of a very bright area in the center of Berlin near Berlin Museumsinsel (stop 1) and a relatively dark area south of Berlin near a village called Petkus (stop 12). The details of the locations and times can be found in appendix A.

Several other options in the West and North of Berlin were considered and tested as well. However, stops within the city limit were difficult to find along those routes. The southern route was also selected because Berlin has a well-defined city limit along this route because of the former border between West and East Germany. Each stop was selected to be relatively well accessible, but at the same time had to provide an as much as possible unobstructed view to the sky and a minimum of direct illumination. The latter is very crucial for the rural sites and impossible to achieve inside the city limits, thus a compromise had to be elaborated. Measurements in the city were therefore performed mainly within city parks to get a free view to the sky of about 60$^{\circ}$ from zenith angle (or only blocking from buildings or trees up to 30$^{\circ}$ above the horizon, respectively).
\begin{figure}[h]
\centering
\includegraphics[width=0.8\columnwidth]{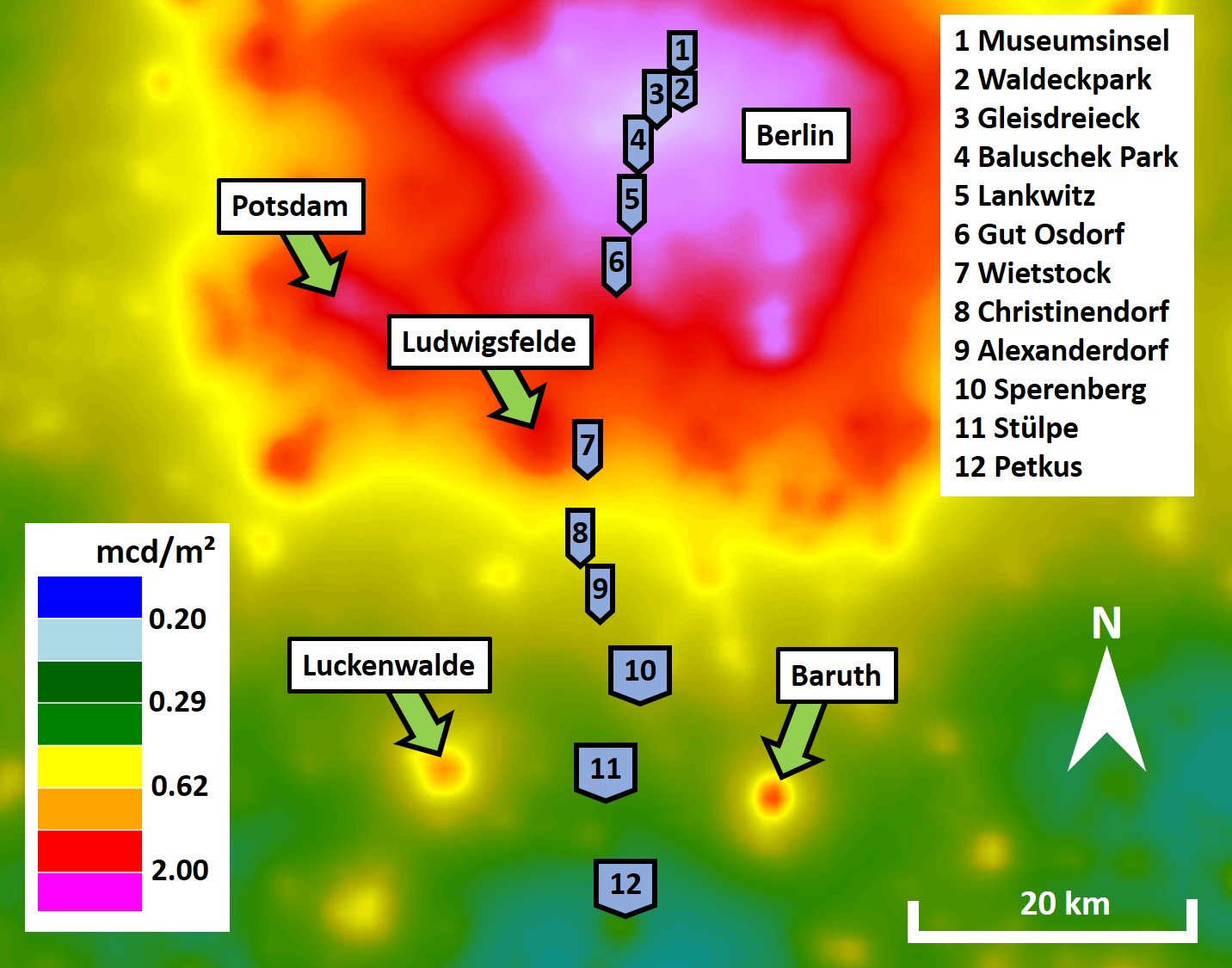}
\caption{Maps of the observation locations in the capital region of Berlin, Germany. (Map from https://www.lightpollutionmap.info, using data from the new world atlas of artificial night sky brightness (NWA) \cite{falchi2016WA})}
\label{map}
\end{figure}

The potential dark rural area was identified from skyglow map data of the new world atlas (NWA) of artificial night sky brightness and then characterized with pilot NSB measurements. The potential stops were first identified from map and skyglow data and the transect was tested during nights with moon up or unstable cloud cover regarding feasibility. In previous work \cite{jechow2017balaguer}, we gained experience performing such transects in a rural context. An overview of the stops of the transect is given in fig. \ref{map} showing a map of Berlin using data from the NWA. Detailed information about each site are listed in the appendix.

\subsection{DSLR camera and software}
We used hemispherical (all-sky) photometry with a commercial DSLR camera and a fisheye lens. This method has several advantages compared to other NSB methods, discussed in a recent review paper \cite{hanel2018measuring}. It provides easy use, quick operation, moderate price and spatially resolved imaging data in three spectral bands. Furthermore, it is applicable for field work \cite{jechow2017balaguer}, even under the challenging conditions like in winter at the Arctic circle \cite{jechow2019snowglow} or on a moving boat in marine waters \cite{jechow2017measuring}. 

In this work, we obtained fisheye images with a Canon EOS 6D commercial DSLR camera. This camera has a full-frame CMOS sensor with 20.2 Megapixel (5496 x 3670 pixels), a built-in GPS sensor, ISO settings from 100 to 25,600 and shutter speed ranges up to 30 s in automatic mode. The camera was operated with a circular fisheye lens (Sigma EX DG with 8 mm focal length) always at full aperture of f/3.5. For more details on how to set up such a camera for night-time light measurements we recommend a recent method paper with a focus on ecology providing valuable information for newcomers \cite{jechow2019beyond}.

To obtain all-sky images, the camera was aligned to image in the horizontal plane with the center of the lens pointing towards the zenith. The camera was aligned within the cardinal coordinates of North and South. If that was not satisfactorily done, the images were rotated using the star tracking option of the processing software for clear skies and manual rotation based on nearby skyglow sources for overcast skies. Shutter speed and ISO settings varied a lot across the transect. We used settings ranging from ISO 3200 and shutter speed of 30s at the darkest observation to ISO 1600 and 0.5s at the brightest location, respectively. The camera obtains multispectral RGB images and stores the files in a raw format.

For image processing we used the commercial "Sky Quality Camera" (SQC) software (Version 1.8, Euromix, Ljubljana, Slovenia). The camera is photometrically calibrated by the software manufacturer, using the green channel of the chip for photometric calibration. The calibration also included correction of optical aberrations like vignetting. 

The software provides the luminance $L_{v}$ for each pixel and it is possible to calculate the illuminance from the luminance data, as well as the CCT from the three color channels. The calibration of SQC is described in more detail in our recent method paper \cite{jechow2019beyond}. The CCT functionality of SQC was tested with several indoor and outdoor light sources (LED lamps, street lamps, moonlight) and verified with a spectroradiometer. We defined the luminance at zenith $L_{v,zen}$ within a circle of 10$^{\circ}$ radius around the zenith.

The cosine corrected illuminance $E_{v,cos}$ in the imaging plane is defined as:
\begin{equation}
E_{v,cos}=\int_{0}^{\pi \over 2}\int_{0}^{2\pi} L_{v,sky}(\theta, \phi) sin\theta cos\theta d\phi d\theta,
\end{equation}
and the scalar illuminance for the imaging hemisphere $E_{v,scal,hem}$ without cosine correction \cite{duriscoe2016photometric} is defined as:
\begin{equation}
E_{v,scal,hem}=\int_{0}^{\pi \over 2}\int_{0}^{2\pi} L_{v,sky}(\theta, \phi) sin\theta d\phi d\theta.
\end{equation}
In the equations, $L_{v,sky}$ is the sky luminance, $\theta$ is the zenith angle and $\phi$ is the azimuth angle. For all-sky images, i.e. when imaging in the horizontal plane, $E_{v, cos}$ is usually termed horizontal illuminance.

For comparison, measurements with a single channel handheld night-sky radiometer, the Sky Quality Meter (SQM, Unihedron, Canada) were performed. The SQM measures the radiance of the night sky $L_{SQM,sky}$ in a spectral band that is close to the photopic and the Johnson V bands, but does not match either precisely. The SQM provides the values in units of mags$_{SQM}$/arcsec$^2$, which can be only approximately converted to a luminance value (please see discussion in \cite{hanel2018measuring}). Both luminance measurements with the DSLR method and the SQM have an error of approximately 10\%." 

\subsection{Night sky brightness and upwelling radiance from night-time remote sensing}
Remotely sensed night-time light data is available from different platforms \cite{levin2019ntl}. The most common spaceborn night-time sensors measure the upwelling radiance in a single  spectral (panchromatic) band, which lacks color information. The first widely applied night-time satellite data set is based on Defense Meteorological Satellite Program (DMSP) Operational Linescan System (OLS) sensor, which collects digital data since 1992. The DMSP/OLS global time series of night time lights with 3 km resolution enabled to study urbanization, socio-economic changes, expansion of road networks and the result of armed conflicts \cite{bennett2017advances}, but suffers mainly from calibration issues. The Day and Night Band (DNB) sensor of the Visible Infrared Imaging Radiometer Suite (VIIRS) on board the Suomi National Polar-orbiting Partnership (Suomi NPP) satellite is tailored for night-time lights \cite{miller2012suomi} and significantly improvement over DMSP/OLS. VIIRS/DNB has a higher spatial resolution of 750 m, is radiometrically calibrated, sensitive to lower light levels and does not saturate in urban areas \cite{elvidge2017viirs}. VIIRS/DNB measures upwelling radiance in a single spectral band from approximately 500 - 900 nm.

Both DMSP/OLS and VIIRS/DNB represent the state of the art in global night-time data that is freely available and have been the basis for a global skyglow model \cite{falchi2016WA}. this model, assumes a certain spatial distribution of the light emitted upwards, also called city emission function \cite{kocifaj2018towards} and normally default scattering parameters. We used the modeled NSB at zenith from the recent version of the world atlas (NWA) \cite{falchi2016WA} extracted from a GIS website\footnote{https://www.lightpollutionmap.info}. The data is based on VIIRS/DNB and is freely available \cite{falchi2016supplement}.

To investigate trends of night-time light emission time series are available from monthly composites. We used “Radiance Light Trends” \footnote{https://lighttrends.lightpollutionmap.info, see report http://www.geoessential.eu/wp-content/uploads/2019/10/GEOEssential-D5.4-Final.pdf}, a web based GIS application to analyse the VIIRS/DNB data for a specific region around the stops of the transect.

For the analysis used in this paper, we used a small region surrounding the observation spots in the city limits of Berlin or the next adjacent settlement for the observation spots outside of the city limits were used. The VIIRS/DNB data were plotted for the longest period available. The web application provides a fit to the data and a factor how much $\%$/year the radiance is increasing or decreasing, respectively. All plots are given in the appendix (figs. \ref{lighttrends1}, \ref{lighttrends2}, \ref{lighttrends3}, \ref{lighttrends4}, \ref{lighttrends5}) and the slope factors are listed in table \ref{table_positions}.

\section{Results}
\subsection{Luminance maps}
\begin{figure}[h]
\centering
\includegraphics[width=\columnwidth]{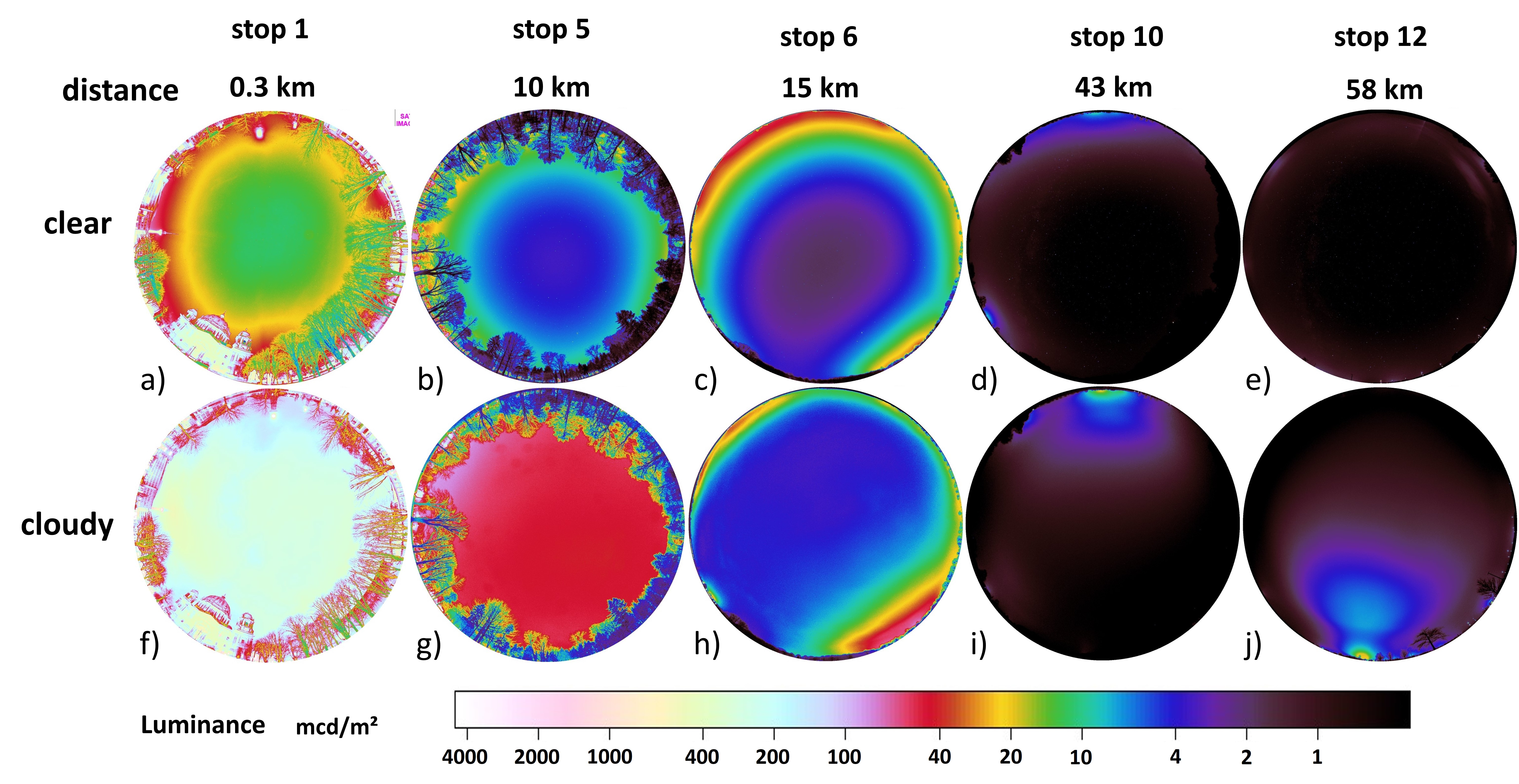}
\caption{Excerpt from the luminance data obtained from the transects in and near Berlin, Germany. The upper row column (a-e) shows data collected during the clear night and the lower row (f-j) shows data collected during the overcast night. The two leftmost columns (a,b,f,g) show data within the city limits, the central column (c,h) data just outside of the city limit and the two rightmost columns (d,e,i,j) data from far outside of the city limits. The distances from the city center are given above the luminance maps. See text for more details and appendix for full data set, exact times and location.}
\label{lum_allsky}
\end{figure}

Figure \ref{lum_allsky} shows an excerpt of the luminance data obtained from the transects in and near Berlin, Germany. The upper row (a-e) shows data collected during the clear night and the lower row (f-j) shows data collected during the overcast night. The two leftmost columns (a,b,f,g) show data obtained within the city limits at stop 1, Berlin Museumsinsel (a,f) and at stop 5, Berlin Gemeindepark Lankwitz (b,g). The central column (c,h) shows data obtained just outside of the city limit of Berlin at stop 6, Gut Osdorf. The two rightmost columns (d,e,i,j) show data obtained far outside of the city limits at stop 10, Sperenberg and at stop 12, Petkus (e,j). The full data set from the 12 (plus 1) stops is given in the appendix.

\begin{figure}[htbp]
\centering
a)\includegraphics[width=0.65\columnwidth]{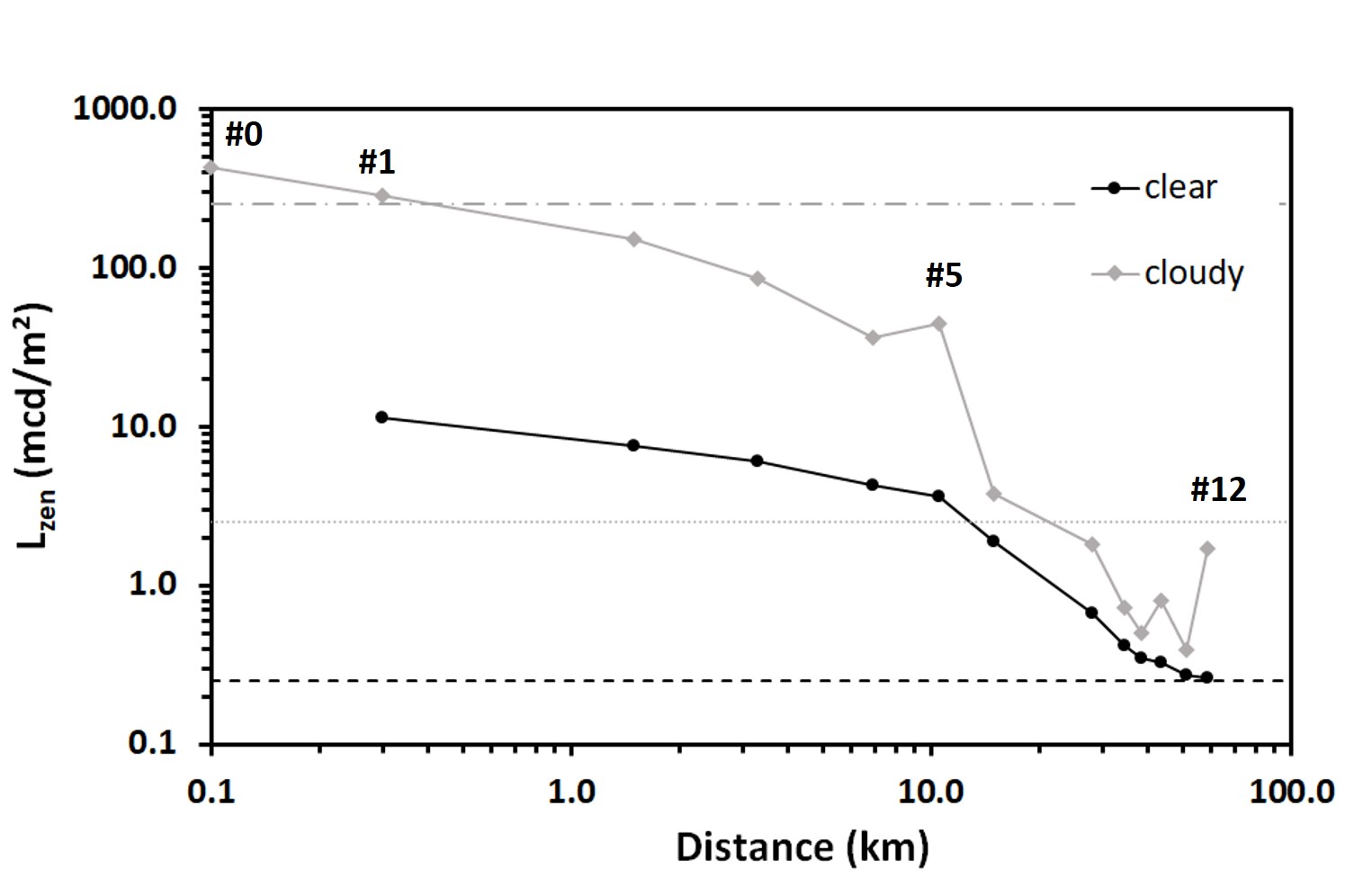}\\
b)\includegraphics[width=0.65\columnwidth]{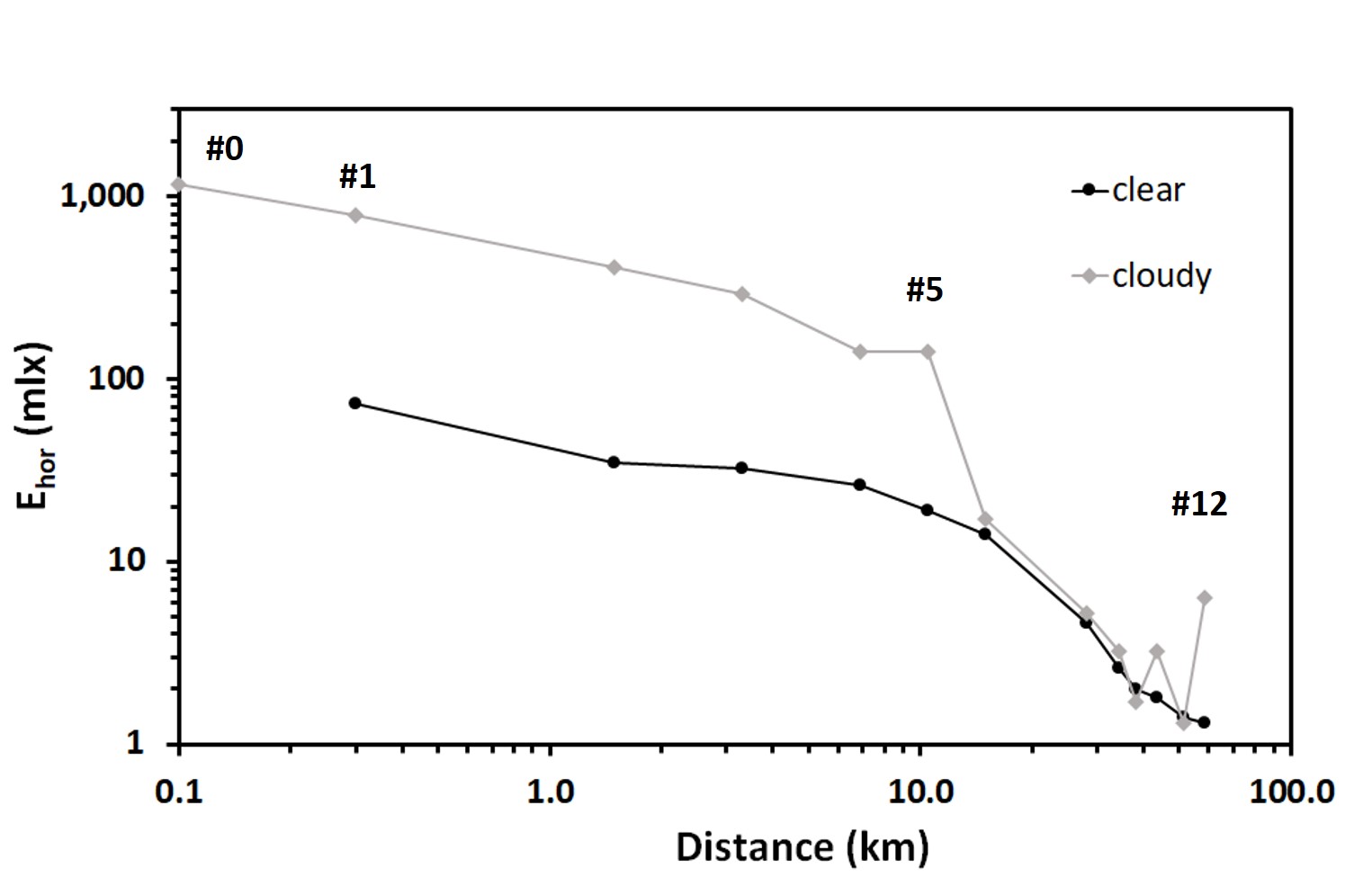}\\
c)\includegraphics[width=0.65\columnwidth]{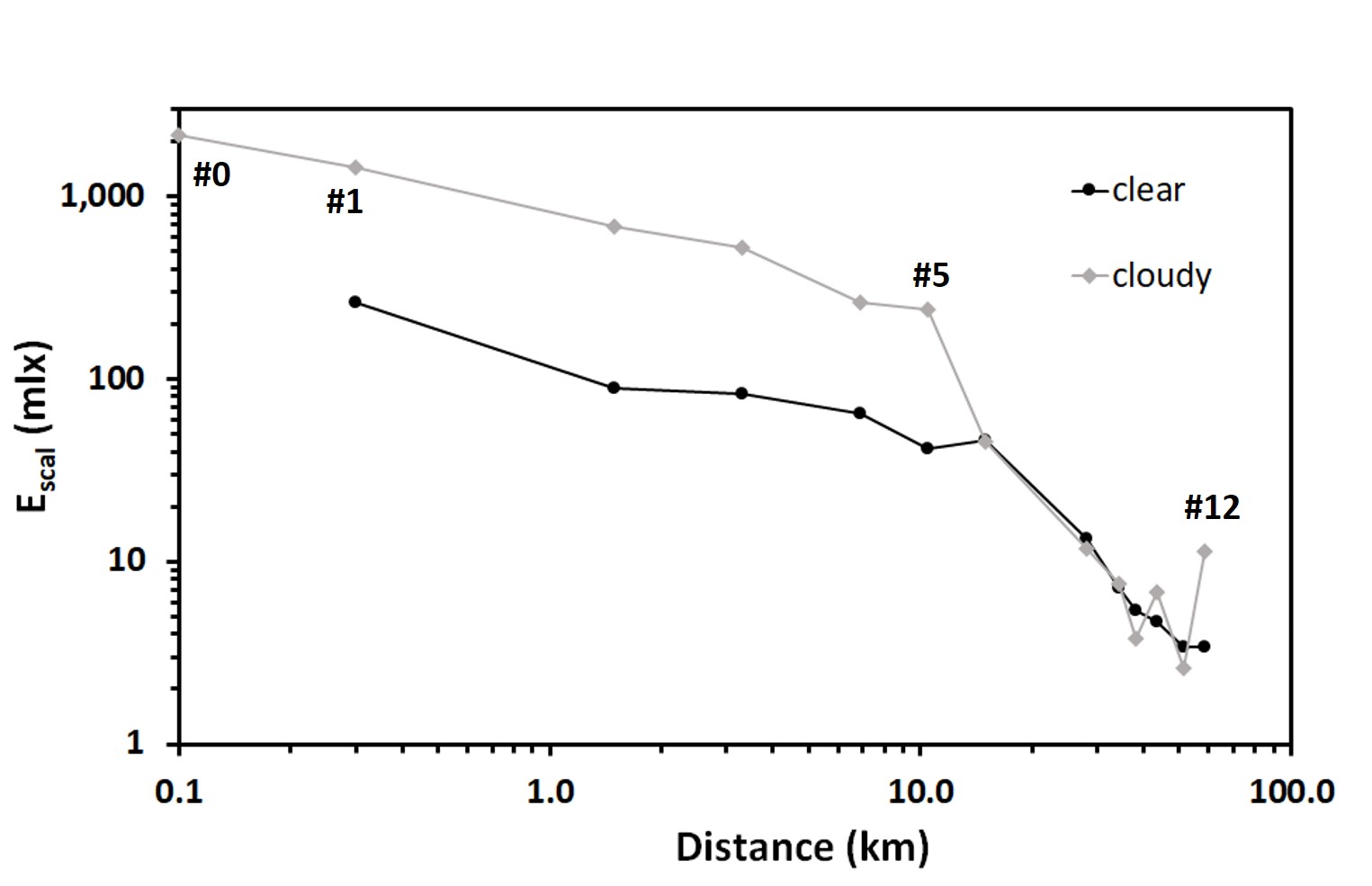}
\caption{Luminance (a), horizontal illuminance (b) and scalar illuminance (c) as a function of distance. The lines in (a) represent the 10 times and 100 times, respectively. The data was extracted from the luminance maps (see methods section for details).}
\label{lum_dist}
\end{figure}

\begin{table}[htbp]
  \centering
  \caption{Photometric data as a function of distance. Zenith luminance, horizontal and scalar illuminance determined with all-sky DSLR photometry. For comparison, values measured with an SQM and values calculated by the skyglow model of the new world atlas (NWA) and the ratios $R_{DLSR/NWA}=L_{v,zen,DSLR}/L_{v,zen,NWA}$ and $R_{DSLR/SQM}=L_{v,zen,DSLR}/L_{v,zen,SQM}$ are given. The error with the DSLR method and the SQM is approximately 10$\%$. $CAF$ - cloud amplification factor. }
  \label{table_dist}
\small
  \begin{tabular}{c}
Zenith Luminance $ L_{v,zen}$ (mcd/m$^2$)  \\
  \end{tabular}
  \begin{tabular}{ccccccccccccc}
	stop nr.&1&2&3&4&5&6&7&8&9&10&11&12\\
	\hline
	\hline
distance&0.3&1.5&3.3&6.9&10.5&15&28&34.5&38.3&43.5&51.5&58.5\\
\hline 
clear&11.3&7.5&6.0&4.3&3.6&1.9&0.67&0.42&0.35&0.33&0.27&0.26\\
overcast&285&151&86.0&36.4&44.8&3.8&1.8&0.72&0.50&0.81&0.39&1.7\\
\hline
$CAF_{cl,zen}$&25.2&20.1&14.3&8.5&12.4&2.0&2.7&1.7&1.4&2.5&1.4&6.5\\
\hline
\hline
NWA&6.00&4.80&3.90&3.00&2.20&1.30&0.55&0.37&0.33&0.29&0.25&0.24\\
$R_{DLSR/NWA}$&1.88&1.56&1.54&1.43&1.64&1.46&1.22&1.14&1.06&1.14&1.08&1.08\\
\hline
\hline
SQM&11.8&6.8&5.2&3.6&3.0&1.7&0.68&0.39&0.36&0.33&0.30&0.27\\
$R_{DSLR/SQM}$&0.95&1.10&1.16&1.20&1.21&1.11&0.98&1.07&0.98&1.01&0.91&0.96\\
\hline
\vspace{0.1cm}
  \end{tabular}

  \begin{tabular}{c}
Horizontal illuminance $ E_{v,hor}$ (mlux)  \\
  \end{tabular}
  \begin{tabular}{ccccccccccccc}
	stop nr.&1&2&3&4&5&6&7&8&9&10&11&12\\
	\hline
	\hline
distance&0.3&1.5&3.3&6.9&10.5&15&28&34.5&38.3&43.5&51.5&58.5\\
\hline 
clear&72.6&34.8&32.4&26.2&19.0&14.0&4.6&2.6&2.0&1.8&1.4&1.3\\
overcast&790&410&290&140&140&17.0&5.2&3.2&1.7&3.2&1.3&6.3\\
\hline
$CAF_{cl,hor}$&10.9&11.8&9.0&5.3&7.4&1.2&1.1&1.2&0.9&1.8&0.9&4.8\\
  \hline
\vspace{0.1cm}
	\end{tabular}

    \begin{tabular}{c}
Scalar illuminance $ E_{v,scal,hem}$ (mlux)  \\
  \end{tabular}
  \begin{tabular}{ccccccccccccc}
	stop nr.&1&2&3&4&5&6&7&8&9&10&11&12\\
	\hline
	\hline
distance&0.3&1.5&3.3&6.9&10.5&15&28&34.5&38.3&43.5&51.5&58.5\\
\hline 
clear&263&88.6&83.3&64.2&41.6&45.8&13.3&7.2&5.4&4.7&3.4&3.4\\
overcast&1440&680.0&520&260&240&45.0&11.7&7.5&3.8&6.8&2.6&11.4\\
\hline
$CAF_{cl,sc}$&5.5&7.7&6.2&4.0&5.8&1.0&0.9&1.0&0.7&1.4&0.8&3.4\\
  \end{tabular}
\end{table}

\subsection{Zenith luminance and illuminance}
From the imaging data, we calculated the zenith luminance $L_{v,zen}$ (in a circle of 10$^{\circ}$ radius), horizontal illuminance $E_{v,hor}$  and scalar illuminance $E_{v,scal}$ at each site for overcast and clear conditions, respectively. The zenith luminance $L_{v,zen}$ ranged from near-natural values of 0.26mcd/m$^2$ for clear skies at the measurement spot furthest away from the city center (stop 12) to a value of more than 1000 times higher than the clear sky reference at the city center (stop 1) for overcast conditions (285mcd/m$^2$). The horizontal illuminance $E_{v,hor}$ ranged from 1.3 mlx for clear skies at the measurement spot furthest away from the city center (stop12) to a value of 790 mlx at the city center (stop1) for overcast conditions. The scalar illuminance $E_{v,scal}$ ranged from 3.4 mlx for clear skies at the measurement spot furthest away from the city center (stop12) to a value of 1440 mlx at the city center (stop1) for overcast conditions.

All of the extracted values from the imaging data set and the comparison with modeled NSB from the NWA (\cite{falchi2016WA}) and SQM are listed in table \ref{table_dist}. The measured SQM values for clear sky conditions match relatively well with the imaging data within the instrument error of 10$\%$. The highest deviation is 21$\%$ at the outskirts of the city limit (stop 4 and 5). The modeled data from the NWA  however, deviates very strongly within the city limits, where the measured value in the city center (stop 1) is almost 90$\%$ higher than the value reported from the NSB modeling. Within the city and just outside of the city (stops 2-6), the measured values are about 50$\%$ higher than values from the model. At stops further outside of the city (stops 7-12), the values match much better reaching the instrumental error range of ca. 10$\%$. However, the model underestimates the NSB for all measurements.

Figure \ref{lum_dist} a) shows the zenith luminance as function of the distance for the clear and the overcast night. The black dashed line represents the reference values for a typical clear night\footnote{please be aware that this definition is not strict and some other authors use a different value of 0.17 mcd/m$^2$ \cite{falchi2016WA, duriscoe2016photometric} but the 0.25 mcd/m$^2$ are sometimes casually called NSU - natural sky unit} (0.25 mcd/m$^2$) \cite{Kyba:2015_isqm}, the grey dotted line ten times this value (2.5 mcd/m$^2$) and the gray dashed dotted line thousand times this value (250 mcd/m$^2$). Figure \ref{lum_dist} b) shows the horizontal illuminance and Fig. \ref{lum_dist} c) shows the scalar illuminance as function of the distance for the clear and the overcast night, respectively.

All three plots show a similar trend: the slope within the city limits is lower than that outside of the city limit and there is a strong drop for the overcast night data for all three measured parameters just outside the city limit (stop 6). The zenith luminance $L_{v,zen}$ for the overcast night never falls below the clear sky value at the individual spots, which should be the case for pristine skies \cite{jechow2019using}. However, the illuminance values for overcast nights at some measurement spots far outside of the city limit actually fall below the clear sky values (stop 9 and 11). It is interesting to note, that at two distant spots (stop 10 and 12), the overcast night data exceeds the clear sky data considerable for all three parameters.
\begin{figure}[htbp]
\centering
\includegraphics[width=0.7\columnwidth]{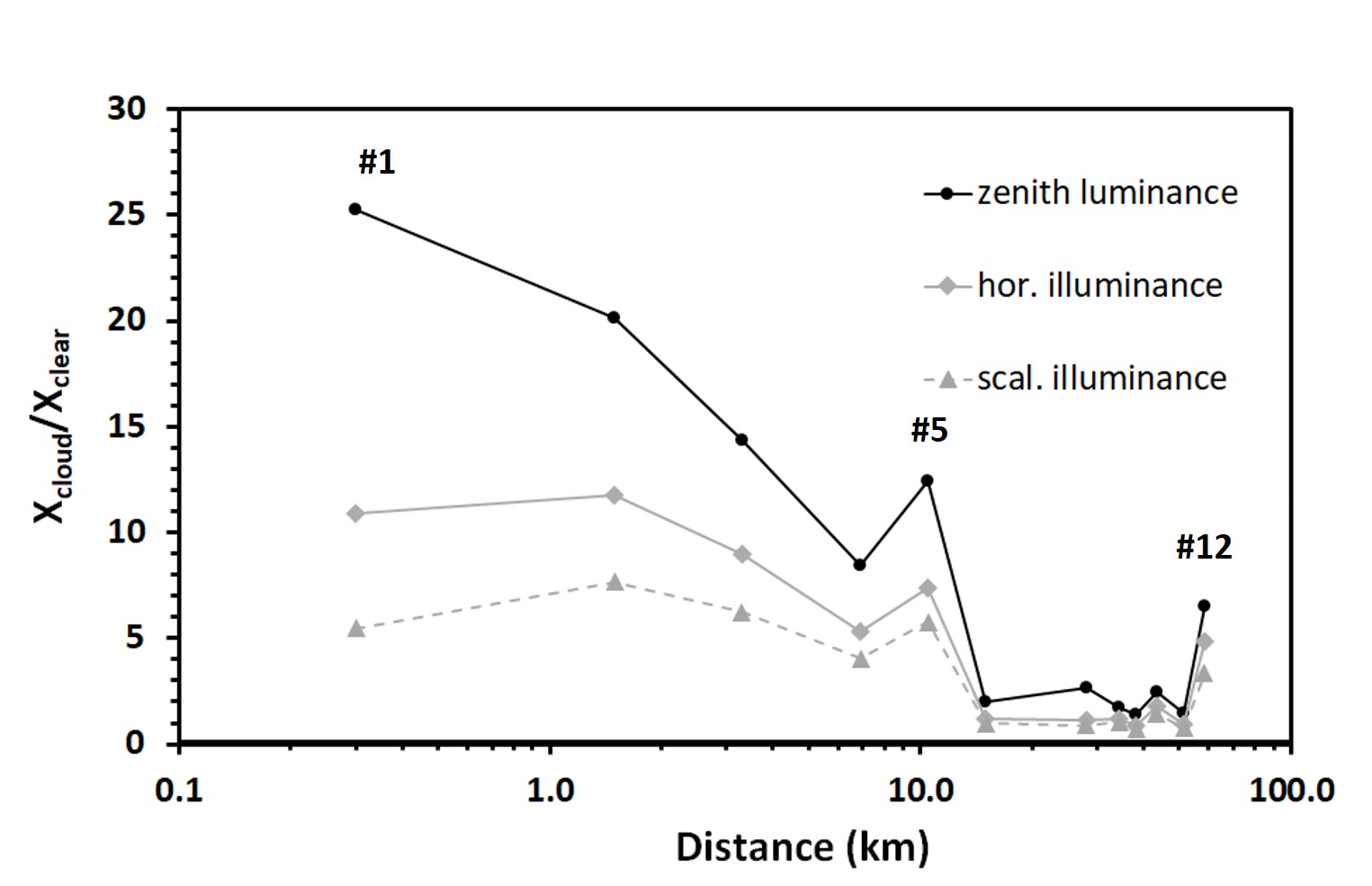}
\caption{Cloud amplification factors $CAF_{cl,x}=X_{cloud}/X_{clear}$ as function of distance for zenith luminance, horizontal illuminance and scalar illuminance.}
\label{amp}
\end{figure}

\subsection{Cloud amplification factors for zenith luminance and illuminance}
From the extracted data, the cloud amplification factors for each measured parameter $CAF_{cl,x}=X_{cloud}/X_{clear}$ can be calculated, where $X$ is a measured parameter for example the zenith luminance $L_{v,zen}$. The individual values are listed in table \ref{table_dist} and are plotted as a function of distance in fig. \ref{amp}.

The cloud amplification factor for zenith luminance $CAF_{cl,zen}$ exceeds a value of 25 in the city center and decreases to around 1.4 outside of the city limit. The cloud amplification factor for horizontal illuminance $CAF_{cl,hor}$ is almost 11 in the city center and very close to unity outside of the city, where even values below 1 were observed which means that clouds darken the sky. A similar trend is observed for scalar illuminance cloud amplification factors $CAF_{cl,sc}$. The lowest value is obtained at stop 9 with $CAF_{cl,sc} = 0.7$. Unexpectedly, at the furthest distant measurement point (stop 12), the cloud amplification factors reach values as high as inside of the city (i.e. stop 4). The variability of cloud amplification factors for the zenith luminance within the city is high, reaching a factor of 3 between city center (stop 1, $CAF_{cl,zen} = 25.2$) and 7 km distance to the city center (stop 4, $CAF_{cl,zen} = 8.5$).

\begin{figure}[htbp]
\centering
a)\includegraphics[width=0.7\columnwidth]{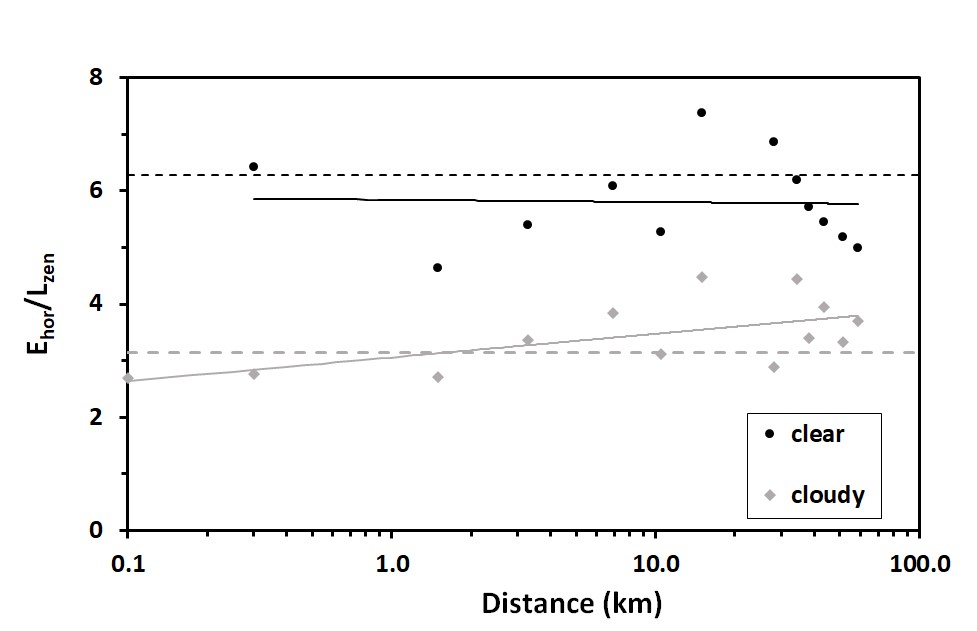}
b)\includegraphics[width=0.7\columnwidth]{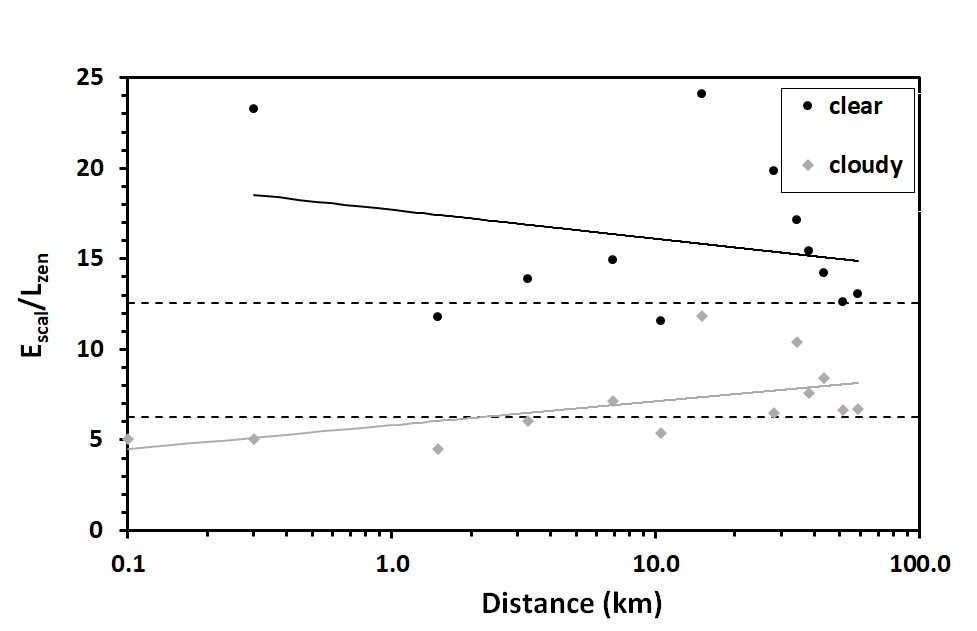}
\caption{Ratio between zenith luminance and illuminance as function of distance for (a) horizontal illuminance and (b) scalar illuminance. We propose to name this ratio the Posch-ratio - $PR$. The solid lines are fits through the data and the dashed lines are multiples of $\pi$.}
\label{Posch}
\end{figure}

\subsection{Ratio between zenith luminance and illuminance}
With the imaging method used in this work, it is possible to extract zenith luminance and illuminance simultaneously. However, the most commonly used devices for NSB monitoring are single channel devices like the SQM, that only measure at zenith. In other fields like for example ecology or chronobiology, the illuminance is an important parameter to know \cite{souman2018spectral, hale2015ecological, grubisic2019light}, allowing to estimate the impact of LP on flora and fauna. Recently, Kocifaj, Posch and Solano-Lamphar \cite{kocifaj2015zenith} have pointed out the importance of this issue and modeled different NSB distributions and different ground albedo scenarios.

We propose to name the ratio between zenith luminance and illuminance, the Posch Ratio $PR$, after one of the co-authors of \cite{kocifaj2015zenith} Thomas Posch\footnote{Thomas Posch was an astronomer, astrophysicist and philosopher dedicated to light pollution studies in a very broad and applied sense. Our friend Thomas passed away recently at young age.}. Here, we investigate our heterogenic data set and plot $PR$ for horizontal illuminance and scalar illuminance as a function of distance. See fig. \ref{Posch} a) for horizontal and b) for scalar illuminance ratios.

For the horizontal illuminance, the factor $PR$ is:
\begin{equation}
PR_{hor} = E_{v,hor} / L_{v,zen},
\end{equation}
and for scalar illuminance:
\begin{equation}
PR_{scal} = E_{v,scal} / L_{v,zen},
\end{equation}
respectively. We find, that despite the very different situations (inner city, outside of the city, direct light etc.) we see the values for $PR_{hor}$ are close to $\pi$ for overcast skies and are close to $2\pi$ for clear skies. Thus, the commonly used approximation $E_{v,hor} \approx \pi L_{v,zen}$ should be only used for overcast skies and $E_{v,hor} \approx 2 \pi L_{v,zen}$ for clear skies. However, we want to point out that this approximation should be only used when no all-sky data is available. We of course prefer the full all-sky data from e.g. imaging tools such as used in this work. For $PR_{scal}$ are close to $2\pi$ for overcast skies and are near or above $4\pi$ for clear skies.

\subsection{Correlated color temperature (CCT)}
\begin{figure}[ht]
\centering
\includegraphics[width=\columnwidth]{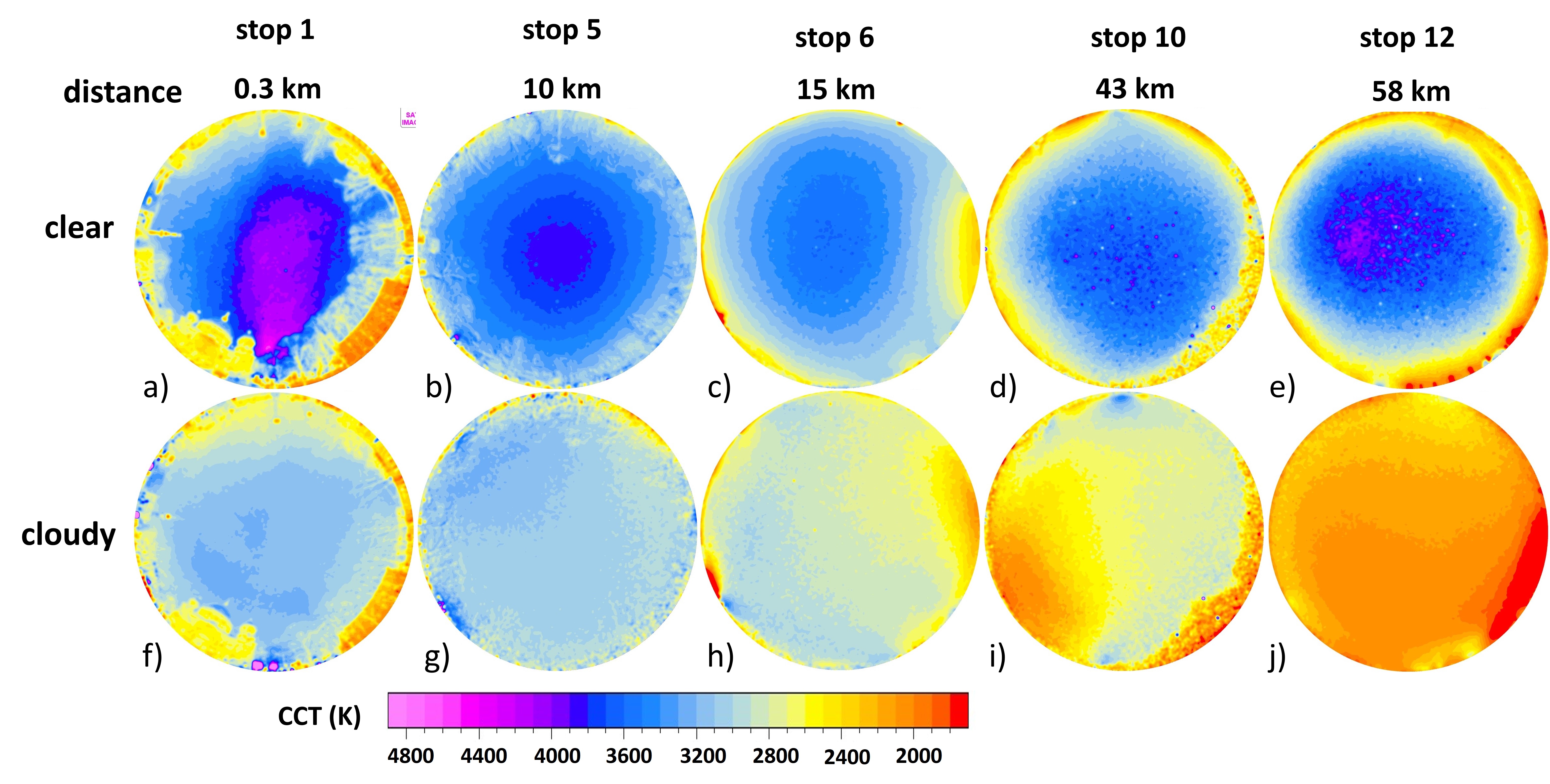}
\caption{Excerpt from the CCT data obtained from the transects in and near Berlin, Germany. The upper row column (a-e) shows data collected during the clear night and the lower row (f-j) shows data collected during the overcast night. The two leftmost columns (a,b,f,g) show data within the city limits, the central column (c,h) data just outside of the city limit and the two rightmost columns (d,e,i,j) data from far outside of the city limits. The distances from the city center are given above the CCT maps. See appendix for full data set and details.}
\label{cct_allsky}
\end{figure}

A big advantage of the DSLR imaging method is the color information in the multispectral (RGB) data. The software SQC allows to produce CCT maps and provides analysis of CCT for different sectors of the imaging data. A part of the CCT maps calculated from the imaging data for clear and overcast conditions for the same stops as used in fig \ref{lum_allsky} are shown in fig. \ref{cct_allsky}.

From the CCT maps it is obvious, that the CCT is highest at zenith for all clear sky measurements (see upper row in fig. \ref{cct_allsky}). For the overcast measurements this changes and some measurements have highest CCT near the horizon. This can be seen for example for stop 6 in fig. \ref{cct_allsky} h) or fig. \ref{stop6} h) in the appendix. Other examples are stop 4 (fig. \ref{stop4}), or stop 5 (fig. \ref{stop5}).

Furthermore, other sky color changes from clear to overcast conditions can be extracted from the CCT maps. For example there is a strong red shift at stop 8 (see fig. \ref{stop8} g) towards the South (lower part of the image), which is probably from the wind energy park. In this example it also is obvious, that the part of the sky with the strongest CCT signature (i.e. CCT difference from background) does not necessarily have to be the brightest part. The brighter sections of the sky for overcast conditions at stop 8 are towards the North or the West, as can be seen in fig. \ref{stop8} d) upper part and right part of luminance map. A similar situation is also visible in Fig. \ref{cct_allsky} i) when compare to the luminance data in Fig. \ref{lum_allsky} i).

\begin{figure}[h]
\centering
a)\includegraphics[width=0.7\columnwidth]{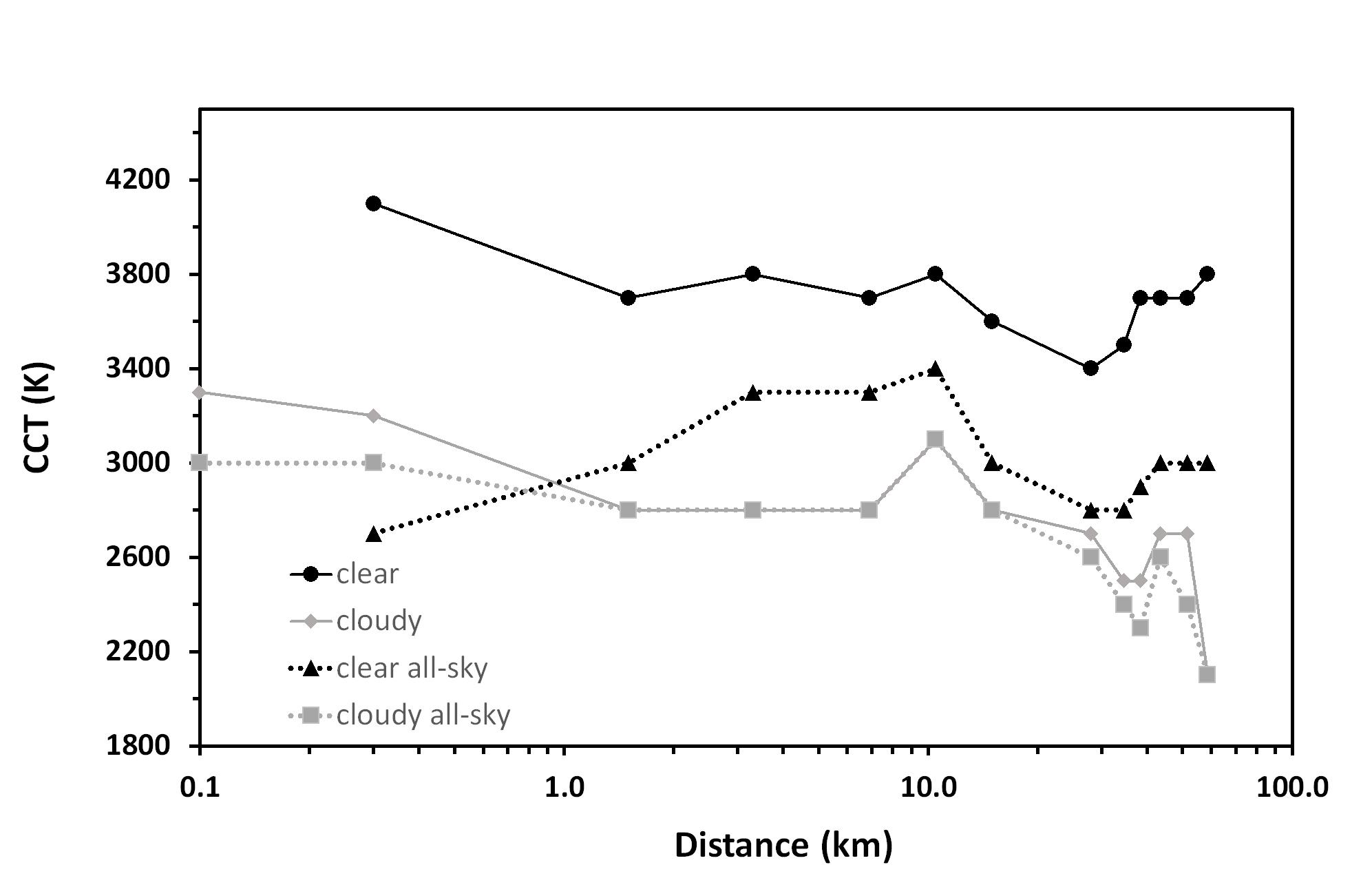}
b)\includegraphics[width=0.7\columnwidth]{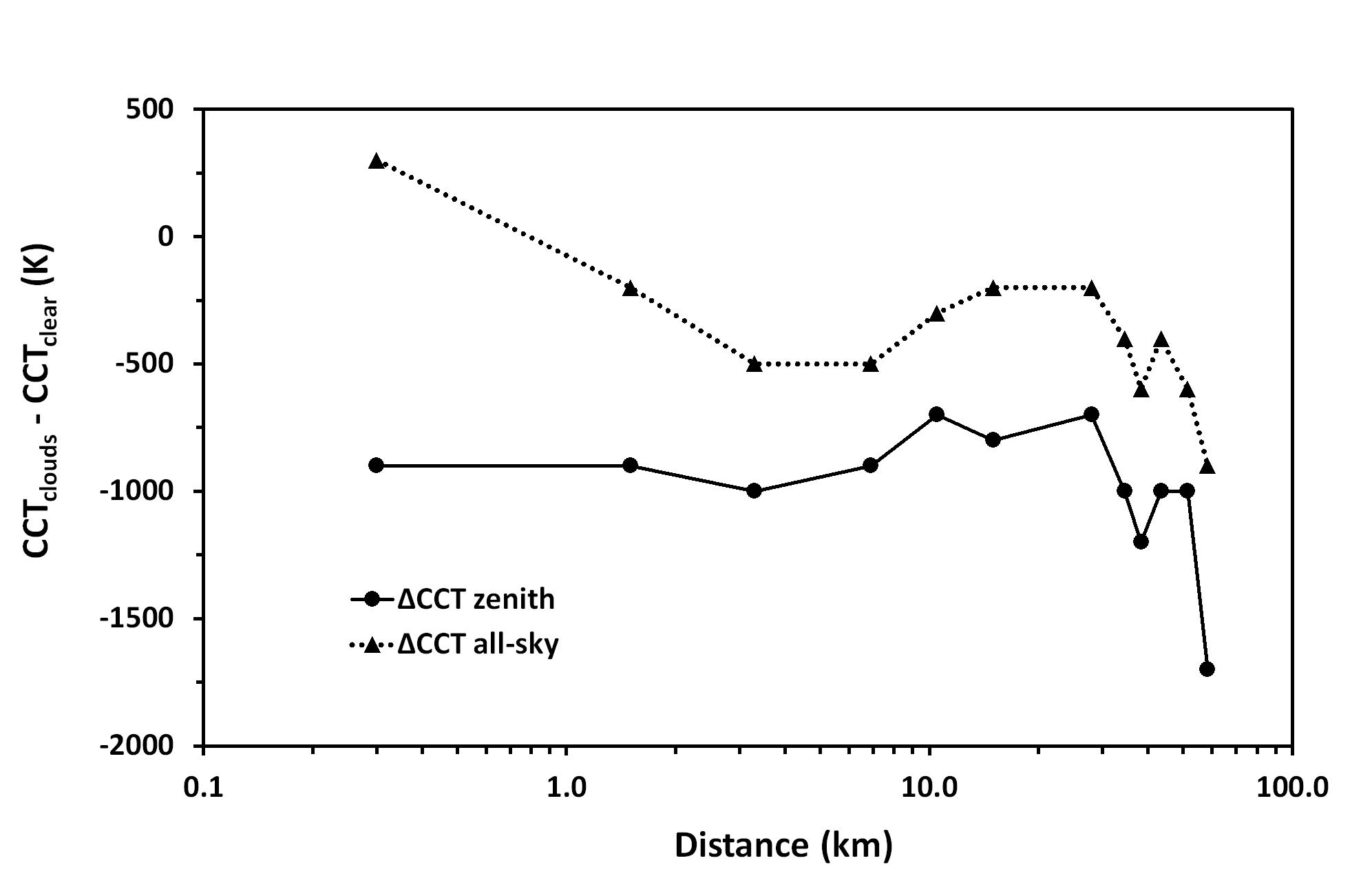}
\caption{(a) CCT as a function of distance for clear and overcast night for zenith and all-sky data and (b) difference in CCT between clear and overcast night for zenith and all-sky data.}
\label{CCT_img}
\end{figure}

The individual values extracted for the zenith (again for a circle with 10$^{\circ}$ radius) and for the full all-sky data (scalar) are listed in table \ref{table_cct}. For the clear sky, the CCT values at zenith range from 3400K to 4100K, with the highest CCT measured in the city center (stop 1, fig\ref{cct_allsky} a). To our knowledge, a complete CCT analysis of pristine skies does not exist. However, we measured values near 4000 K in very rural places (Kasachstan unpublished, South Africa \cite{jechow2019using}) with Milky Way near zenith and low airglow. For overcast sky, the CCT values at zenith are always at lower compared to the clear sky. Values ranged between 2100K and 3200K, with again the highest CCT being measured in the city center. The difference between clear and overcast thus is between 700K and 1700K. However, most values for the CCT difference range between 700K and 1000K and the extreme value of 1700K is only observed at the measurement point furthest away from the city center (stop12). The CCT values as a function of distance for both overcast conditions are shown in fig. \ref{CCT_img} a) and the difference in fig. \ref{CCT_img} b). For the CCT, the city limit does not appear very distinct in the distance plots. When investigating the CCT all-sky data, it is obvious that the difference between clear and cloudy all-sky CCT data is lower than for the zenith CCT data. Furthermore, it is interesting to note that the difference in CCT value between zenith and all-sky data differs less for the cloudy night compared to the clear night. Further investigation of the CCT data near the horizon and of sectors would be possible (see for example \cite{jechow2018tracking} were this was applied to luminance data).

\begin{table}[htbp]
  \centering
  \caption{The CCT determined at zenith and for the full all-sky data (scalar) at each stop (or distance to the city center, respectively) for clear and overcast conditions, the difference $D_{CCT,cl}=CCT_{clouds}-CCT_{clear}$ is also provided.}
  \label{table_cct}
\small
  \begin{tabular}{c}
Correlated color temperature CCT (K) \\
 \end{tabular}
  \begin{tabular}{ccccccccccccc}
stop nr.&1&2&3&4&5&6&7&8&9&10&11&12\\
\hline
distance&0.3&1.5&3.3&6.9&10.5&15&28&34.5&38.3&43.5&51.5&58.5\\
\hline 
&&&&&&zenith&&&&&&\\
clear&4100&3700&3800&3700&3800&3600&3400&3500&3700&3700&3700&3800\\
overcast&3200&2800&2800&2800&3100&2800&2700&2500&2500&2700&2700&2100\\
\hline
$D_{CCT,cl}$&-900&-900&-1000&-900&-700&-800&-700&-1000&-1200&-1000&-1000&-1700\\

&&&&&&scalar&&&&&&\\
clear&2700&3000&3300&3300&3400&3000&2800&2800&2900&3000&3000&3000\\
overcast&3000&2800&2800&2800&3100&2800&2600&2400&2300&2600&2400&2100\\
\hline
$D_{CCT,cl}$&+300&-200&-500&-500&-300&-200&-200&-400&-600&-400&-600&-900\\

  \end{tabular}
\end{table}

\pagebreak
\section{Discussion}
The data set provides much information about the brightness and the color of the night sky across a large gradient.

\subsection{High luminance and illumination values, cloud amplification factors}
The highest zenith luminance for overcast sky was 430mcd/m$^2$ at the very city center (stop 0, Alexanderplatz, fig \ref{stop0}). This is 1700 times higher than a commonly used clear sky value of 0.25mcd/$^2$ and more than 6000 times brighter than the lowest zenith luminance we measured so far for an overcast night \cite{jechow2019snowglow}. A high illuminance of 1.1lx was measured at this position, which is almost 4 times as bright as the maximum full moon illuminance of 0.3lx possible on Earth \cite{hanel2018measuring} (which is more a hypothetical value as it almost never occurs and a typical summer full moon in Berlin is rather around 0.1lx). For this spot however, no clear sky reference measurement was performed.

For the data set with clear and overcast night data, the maximum change was observed in the city center (stop 1, fig. \ref{stop1}). At this measurement spot, the zenith luminance $L_{v,zen}$ increased from 11.3 to 285mcd/m$^2$ and the horizontal illuminance $E_{v,hor}$ increased from ca. 0.073 to 0.79lx, again exceeding full moon light levels and reaching amplification factors of $CAF_{cl,zen} \approx 25$ and $CAF_{cl,hor} \approx 11$, respectively.

There was a gradient found within the city (stops 1-5, fig. \ref{stop1}, \ref{stop2}, \ref{stop3}, \ref{stop4}, \ref{stop5}) for the cloud amplification within the city. Outside of the city the cloud amplification factors decreased significantly compared to the values observed within the city (stops 6-11, \ref{stop6} ,\ref{stop7}, \ref{stop8}, \ref{stop9}, \ref{stop10},\ref{stop11}), apart from some local peculiarities due to local light pollution sources (stop 12, fig. \ref{stop12} see panel d).

Our cloud amplification values are higher than what was reported by a skyglow model for overcast conditions \cite{kocifaj2014quantitative}. This model also predicted that the peak amplification is not reached in the city center. We did not observe this but we also have to point out that the data was not obtained simultaneously and with limited spatial resolution within the city. However, for low (stratus) clouds a strong decrease at the city limit was predicted by this model \cite{kocifaj2014quantitative}, which agrees with our experimental observation. In our observations, low altitude stratus or stratocumulus clouds were present. The cloud base height was estimated to be on the order of 300 m based on the visibility of the iconic Berlin TV Tower (see appendix fig. \ref{stop0}.

\subsection{Color shift}
We observed an overall color shift from clear to overcast condition. All CCT values at zenith and almost all for the all-sky data for overcast conditions were lower than for clear skies, which means that the colors have shifted towards the red. This is in accordance with earlier observations using SQMs and color filters \cite{Kyba:2012_mssqm}. From the CCT maps also spatial information could be derived. We compared CCT at zenith with the full all-sky CCT data and found that the difference between clear and cloudy conditions was highest near zenith. Furthermore, the CCT data was more homogeneous for the overcast conditions were the zenith data differed less from the full all-sky data. Spatially resolved color information is a clear advantage of DSLR photometry over other methods, which was not fully exploited in this analysis. Future work could focus on investigating the CCT as a function of distance from a city by analyzing a specific part of the sky only \cite{jechow2018tracking, jechow2019using}. 

\subsection{Identifying local light pollution sources during overcast conditions}
There were strong local influences on the NSB, particularly for the overcast nights. This occurred most obviously at the measurement point at the largest distance from the city center (stop 12, see appendix figure \ref{stop12} for the full dataset). While the expectation was that the NSB will decrease further with the presence of clouds \cite{Jechow2016,jechow2019using}, a strong increase of NSB was observed. The advantage of imaging devices is that the source can be localized, while with single channel devices this is not possible. For this particular spot, a church at 800 m distance in the village of Petkus could be identified at 172$^{\circ}$ azimuth.

We have observed the unexpectedly high impact of local LP in previous work \cite{jechow2019using}. We conclude that such local LP sources that over-proportionally emit upwards or even horizontally can be easier detected during overcast conditions than with clear skies. This could become relevant for dark sky parks in the future. Another example for such local LP was observed at stop 7 in our dataset (appendix fig. \ref{stop7}). There, a church light is clearly visible within the overcast night luminance map. The church light contributes to increased NSB at zenith, but not to increased illuminance significantly.

\subsection{Deviation between measurements and modeled NSB from NWA}
In previous work, we usually found good agreement between ground based NSB measurements and modeled data from the NWA \cite{jechow2018tracking, jechow2017balaguer}. However, in the present data set we observed a deviation of almost 90$\%$ for the stop in the city center and about 50$\%$ for the other stops within the city limit, while there was better agreement for the stops far outside of the city. There are several potential reasons for this deviation.

Firstly, the aerosol content can vary dramatically between city and countryside due to air pollution. For the clear day before the first measurement night, the aerosol optical depth in Berlin (Aeromet FUB) was 0.16 and at a station about 60 km outside of Berlin (Aeromet Lindenberg) it was much less 0.08. While the FUB station was close to the transect, the Lindenberg station was not on the route of the transect. Ges et al. \cite{ges2018light} performed a transect out into the Mediterranean sea near Barcelona, Spain, with SQMs and found that when the aerosol optical depth is close to that used in the NWA, then the values matched remarkably. However, when they had different aerosol optical depth, the deviation was substantial. 

Secondly an explanation can be that the spatial emission is not homogeneous within the city and that the ratio between direct uplight and light propagating under an angle is varying a lot. The horizontal component of ALAN cannot be easily detected directly by Earth observation satellites like the VIIRS/DNB sensor. However, skyglow models account for that by using a specific spatial distribution that is called the city emission function \cite{kocifaj2018towards}. In the NWA skyglow model, a global average city emission function was assumed which could deviate from the Berlin spatial emission.  

Thirdly, it cannot be ruled out that there is a small amount of straylight entering the wide field optics, particularly in the very center of the city. In this work, we took care and tried to minimize the influence of direct lamps in our measurements. We also checked the effect of stray light in previous work \cite{jechow2017balaguer} by adding an aperture and found no such influence.  

A further source of error is the different spectral sensitivity of the instruments \cite{bouroussis2018effect}. The VIIRS/DNB sensor lacks sensitivity for short wavelength radiation, and the SQM and the DSLR spectral bands have also a mismatch to each other and to the photometric band. The CCT in the city center was considerably higher than towards the edge and outside of the city (also for overcast nights when mainly ALAN is detected). Still, this alone would probably not cause the large deviations observed and the CCT between stops 3 and 8 were rather constant.

Another possibility is that the lighting situation has changed dramatically. The NWA NSB is based on VIIRS/DNB data from 2015 while our transect was performed 2 years later. First of all, it is very unlikely that this can explain a 90$\%$ deviation. Furthermore, the time series are available now for longer periods. We analyzed the VIIRS data from 2012 until 2018 for some regions around the measurement stops using the intuitive webapp lighttrends. The radiance was relatively stable for Berlin as a whole and for most places and changed between +3$\%/a$ and -3$\%/a$ for some stops (table \ref{table_trends} in appendix). This would not explain the deviation that we measured. Furthermore, the lighttrends data should be treated with caution when a small number of pixels is selected because variations from sensor pointing and seasonal variations (e.g. snowcover) can play a role. Please see the methods section for a more detailed description and the appendix for the time series plots in figs. \ref{lighttrends1}-\ref{lighttrends5} and the time series data for each stop in table \ref{table_trends}.

Taking all of the above into account, we assume that the deviation is caused by a combination of many factors together: air pollution, spectral mismatch of the sensors (specifically for high CCTs) and mismatch between city emission function assumed in models and actual spatial distribution. 

\subsection{Weakness of the present data set and possible future improvements}
While many information could be extracted, the present data set is unfortunately not perfect and there is room for improvement.

The first missing point is the clear night data from Alexanderplatz (stop 0), which only caught our attention during the following overcast night. There was no further clear night following the measurement nights.

One weakness of this data set is that the transect took relatively long time of about 4 hours. Thus, changes in switching off of lights or small atmospheric changes cannot be tackled. However, in principle this would be possible with differential photometry as demonstrated for WFF Earth Hour \cite{jechow2019observing} or switch offs of ornamental lights \cite{jechow2018tracking}. For future work it would be much better to either have an instrument that measures a time series at each point. Because this is challenging regarding material and personnel, a compromise would be to have at least one camera stationed in the city center and one far outside of the city, both measuring time series and a third camera measuring at multiple points. 

Another weakness is that there are multiple small and medium sized towns (e.g. Ludwigsfelde) and several villages along the route and nearby (e.g. Baruth, Luckenwalde, Potsdam). To avoid this, an isolated town has to be found. However, this is very challenging. For example, Biggs et al. did a transect with SQMs near Perth, Australia, the most isolated large city in the world but still had the same problem with smaller towns \cite{Biggs:2012}. As already mentioned, Ges et al. \cite{ges2018light} did the transect towards the sea but were missing on the city center data. Las Vegas, Nevada, USA could be a good candidate city.

Several other improvements and future research directions are considerable. For example a lighting inventory could be analyzed and the number of lamps, the lamp types and the distance to the measurement points could be investigated. Then a linkage of the spectral and spatial emission to the night sky brightness and color of the night sky would be possible.  Furthermore, the effect of the population density on the night sky brightness could be checked.

\subsection{Comparison with existing work}
City NSB have been measured before. Zamorano et al.  used an SQM mounted on a vehicle to map the NSB around Madrid in a comprehensive campaign \cite{zamorano2016testing}. This produces a much higher density of data points than what is possible with our method. However, the problem is that roads are often illuminated and that it is difficult to disentangle straylight from lamps from a real measurement (which the authors did account for). For example the skyglow at stop 12 (Petkus, fig. \ref{stop12}) during the overcast night would have been difficult to interpret with SQM measurements. Furthermore, the SQM itself has several drawbacks that have been discussed in earlier work \cite{bouroussis2018effect, hanel2018measuring}.

Another interesting work on cloud amplification was also performed in Madrid by Aube et al. using the advanced ASTMON device \cite{aube2016spectral}. This produced very detailed data but was performed at a fixed point. Our results show that there can be a strong gradient within the city limits, so more data at multiple sites would be desirable. However, the advanced ASTMON is designed for stationary operation and not broadly available.

Some work with DSLR photometry exists. We performed a transect in a rural area in Spain \cite{jechow2017balaguer} and Kocifaj et al. \cite{kocifaj2019asymptotic} did observe multiple sites near Vienna, Austria and Bratislava, Slovakia. However, inner city data was not obtained with DSLR photometry before.

\section{Summary and conclusion}
In summary, our data represents the first comprehensive transect of NSB measurements with a multispectral imaging device to (partially) map the skyglow of a city. We observed very high luminance values for clear skies within the city, much higher than predicted by a commonly used global skyglow model. We further measured very high illuminance values both for clear sky (0.07 lx) and overcast sky (1.1 lx). These values are close or exceed typical full moonlight levels at the latitude of Berlin (0.1 lx). We observed a gradient in luminance and illuminance and a difference in cloud amplification between inside or outside of the city. Furthermore, the multispectral data allowed to analyze color changes with clouds. The calculated CCT values at zenith were all lower for overcast skies than for clear skies. The spatially resolved data unraveled some peculiarities of the NSB and color, like local light pollution sources in the rural areas.

We also discus strength and weaknesses of the method and the dataset itself. In conclusion, the method is very well suited for NSB and color mapping in urban and rural areas and with the usage of multiple devices it will be possible to obtain detailed spatio-temporal NSB and CCT maps of human settlements. Such data will be helpful to fine tune local skyglow models that can predict impact on a specific environment much better than global models.

\section*{Acknowledgements}
We thank the LoNNe group 1 for important impulses regarding NSB measurements with digital cameras.

\section*{Funding}
AJ is supported by the ILES project (SAW-2015-IGB-1) and the CONNECT project (SAW-K45/2017) both funded by the Leibniz Association, Germany and by the IGB Leibniz Institute through the Frontiers in Freshwater Science project (IGB Frontiers 2017). CK acknowledges funding from the Helmholtz Association Initiative and Networking Fund under grant ERC-RA-0031 and by the European Union's Horizon 2020 research and innovation programme ERA-PLANET, grant agreement no.~689443, via the GEOEssential project.

\section*{Declaration of interests}
The authors declare that they have no known competing financial interests or personal relationships that could have appeared to influence the work reported in this paper.





\appendix

\section{Full data set}
\begin{figure}[htp]
\centering
\includegraphics[width=\columnwidth]{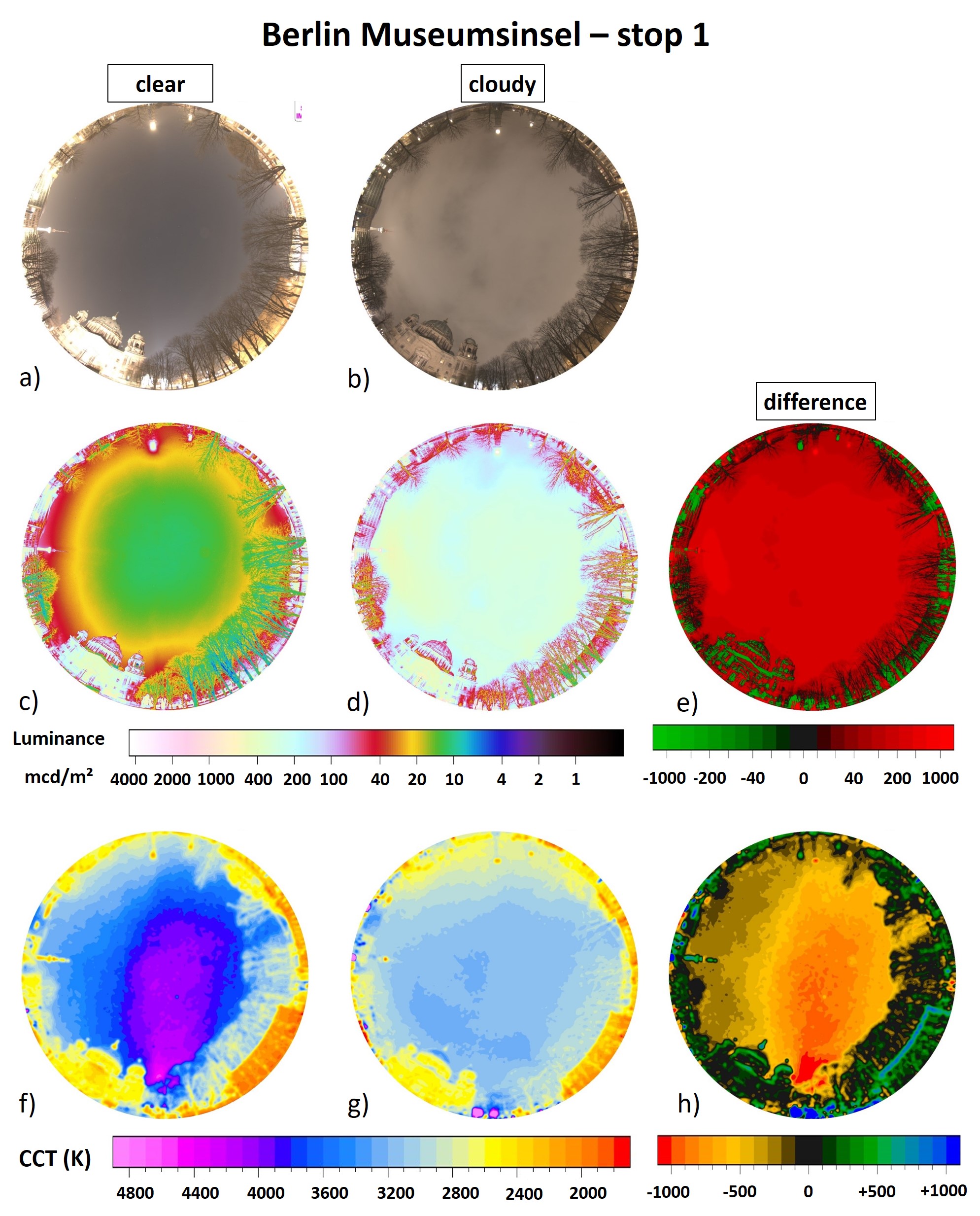}
\caption{All-sky images and calculated imaging data for stop 1 at Berlin, Museumsinsel. The left column shows clear sky data, the middle column overcast sky data and the right hand column subtracted data. a) and b) RGB images; c,d,e luminance; f,g,h CCT.}
\label{stop1}
\end{figure}

\begin{figure}[htp]
\centering
\includegraphics[width=\columnwidth]{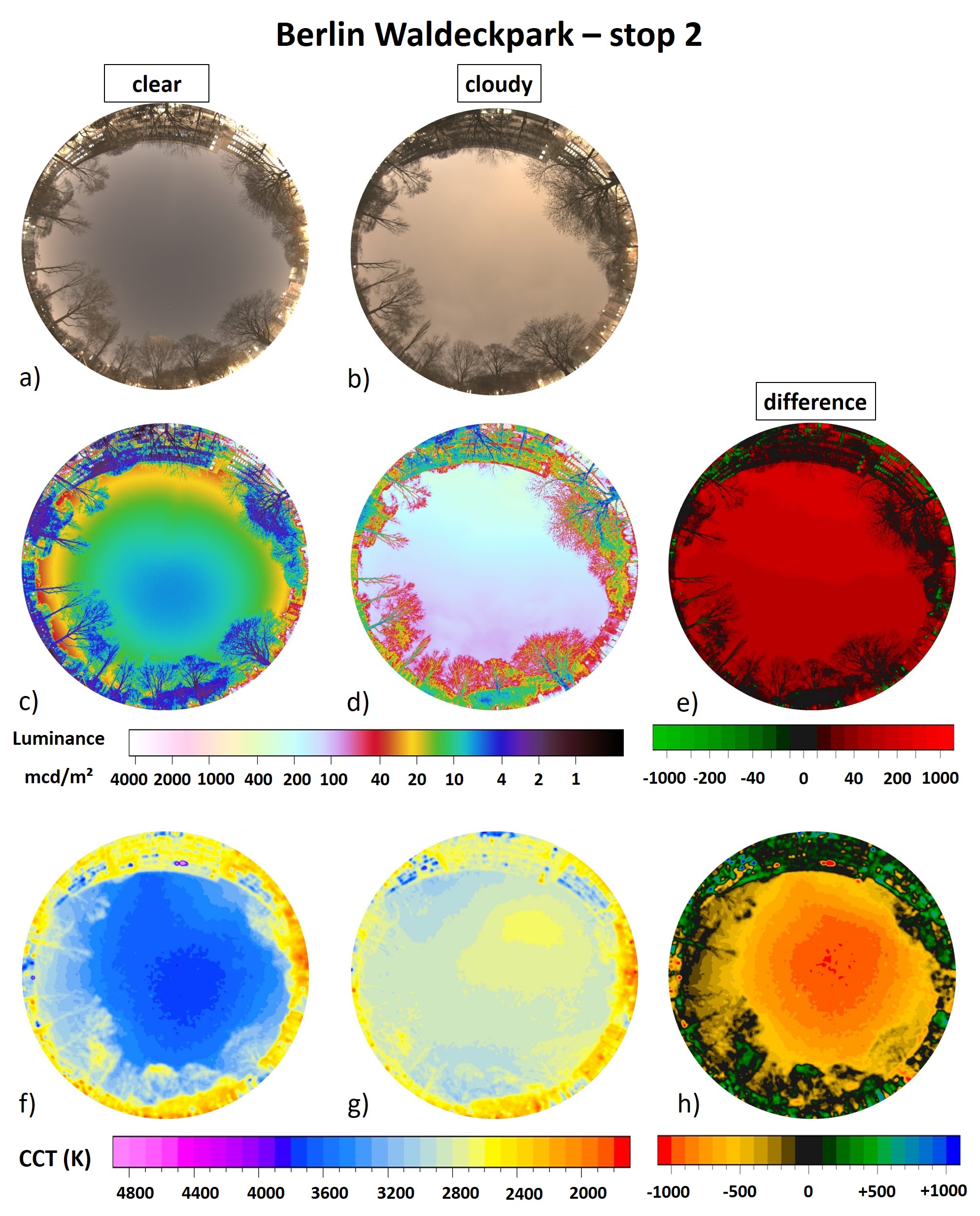}
\caption{All-sky images and calculated imaging data for stop 2 at Berlin, Waldeckpark. The left column shows clear sky data, the middle column overcast sky data and the right hand column subtracted data. a) and b) RGB images; c,d,e luminance; f,g,h CCT.}
\label{stop2}
\end{figure}

\begin{figure}[htp]
\centering
\includegraphics[width=\columnwidth]{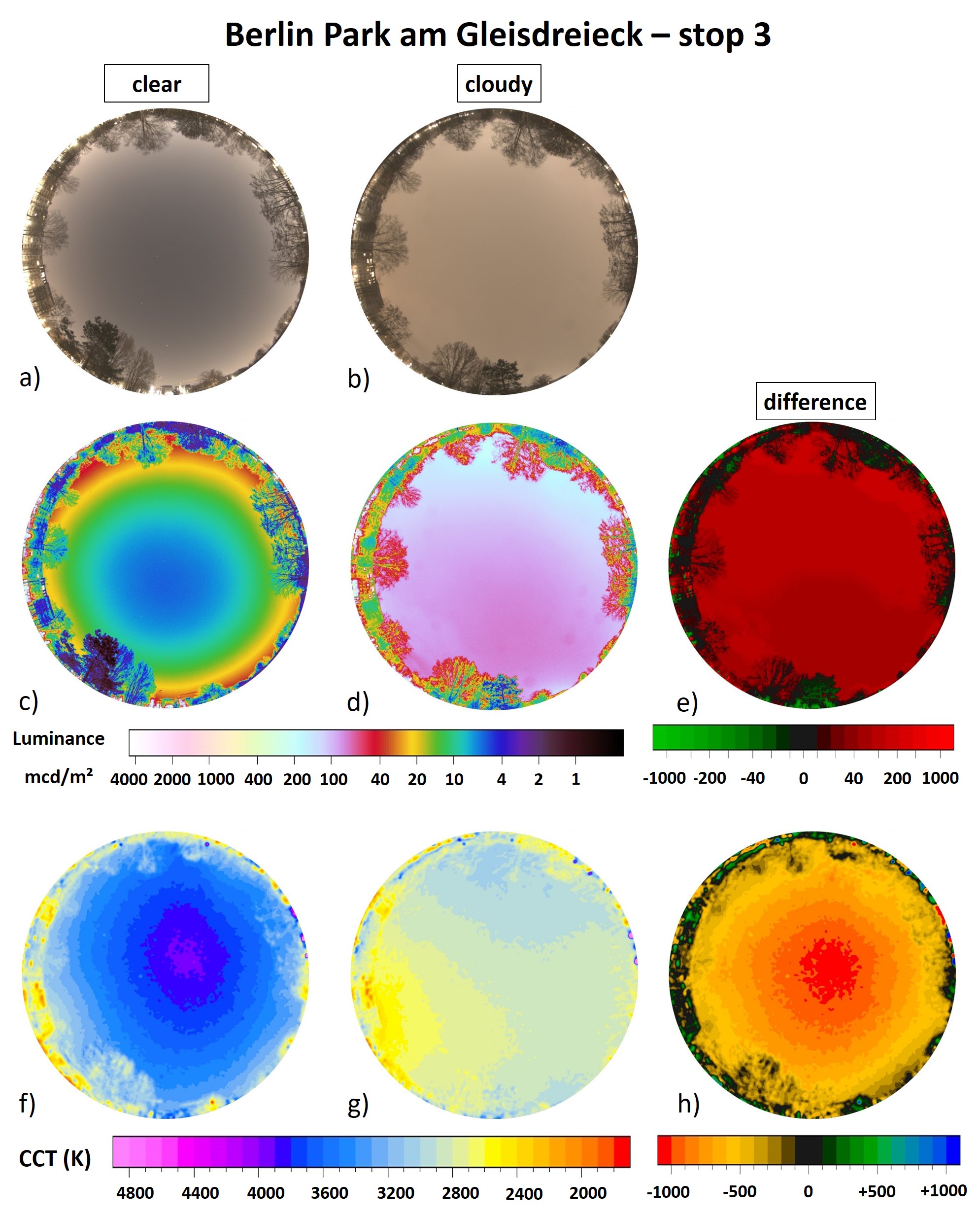}
\caption{All-sky images and calculated imaging data for stop 3 at Berlin, Park am Gleisdreieck. The left column shows clear sky data, the middle column overcast sky data and the right hand column subtracted data. a) and b) RGB images; c,d,e luminance; f,g,h CCT.}
\label{stop3}
\end{figure}

\begin{figure}[htp]
\centering
\includegraphics[width=\columnwidth]{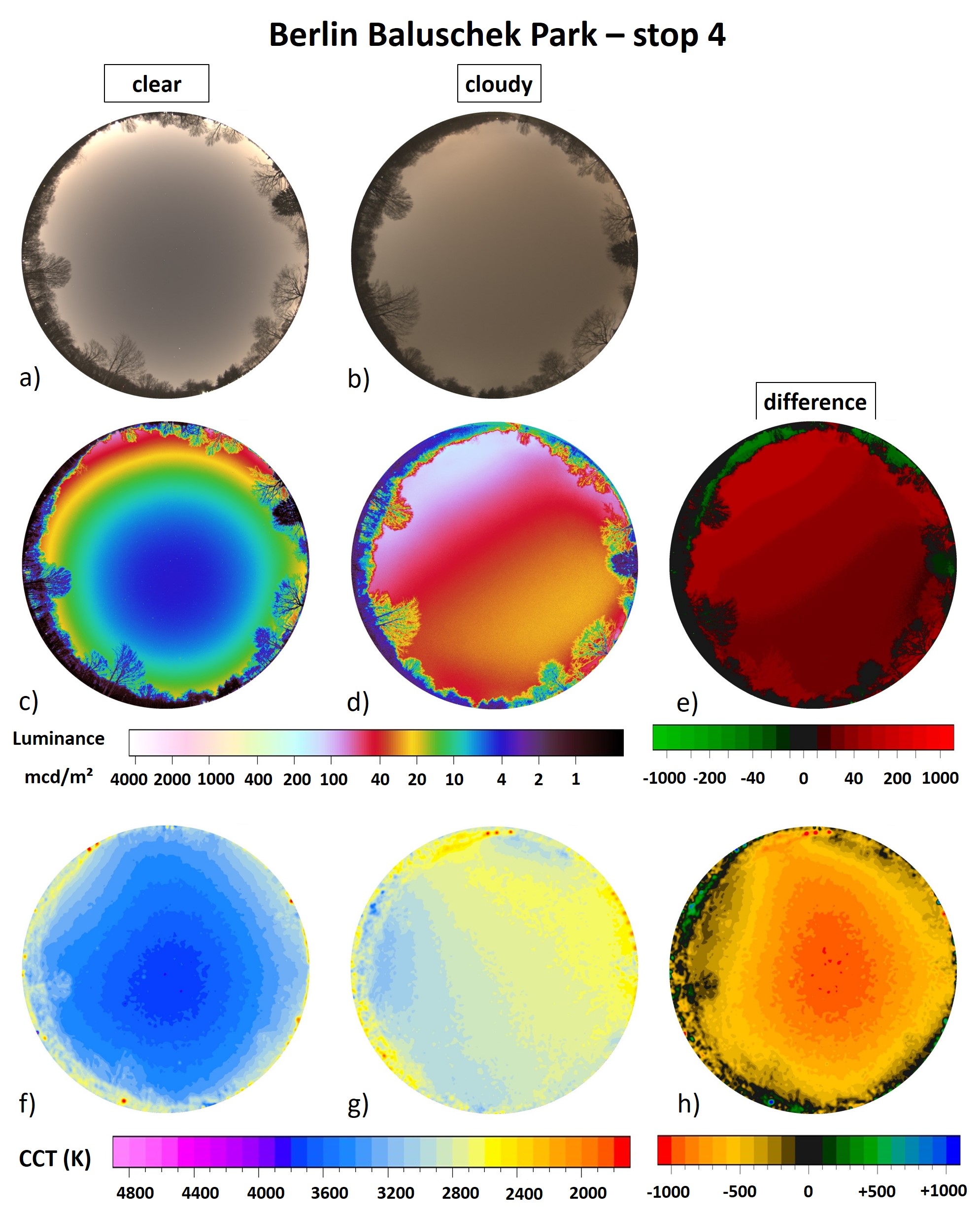}
\caption{All-sky images and calculated imaging data for stop 4 at Berlin, Baluschek Park. The left column shows clear sky data, the middle column overcast sky data and the right hand column subtracted data. a) and b) RGB images; c,d,e luminance; f,g,h CCT.}
\label{stop4}
\end{figure}

\begin{figure}[htp]
\centering
\includegraphics[width=\columnwidth]{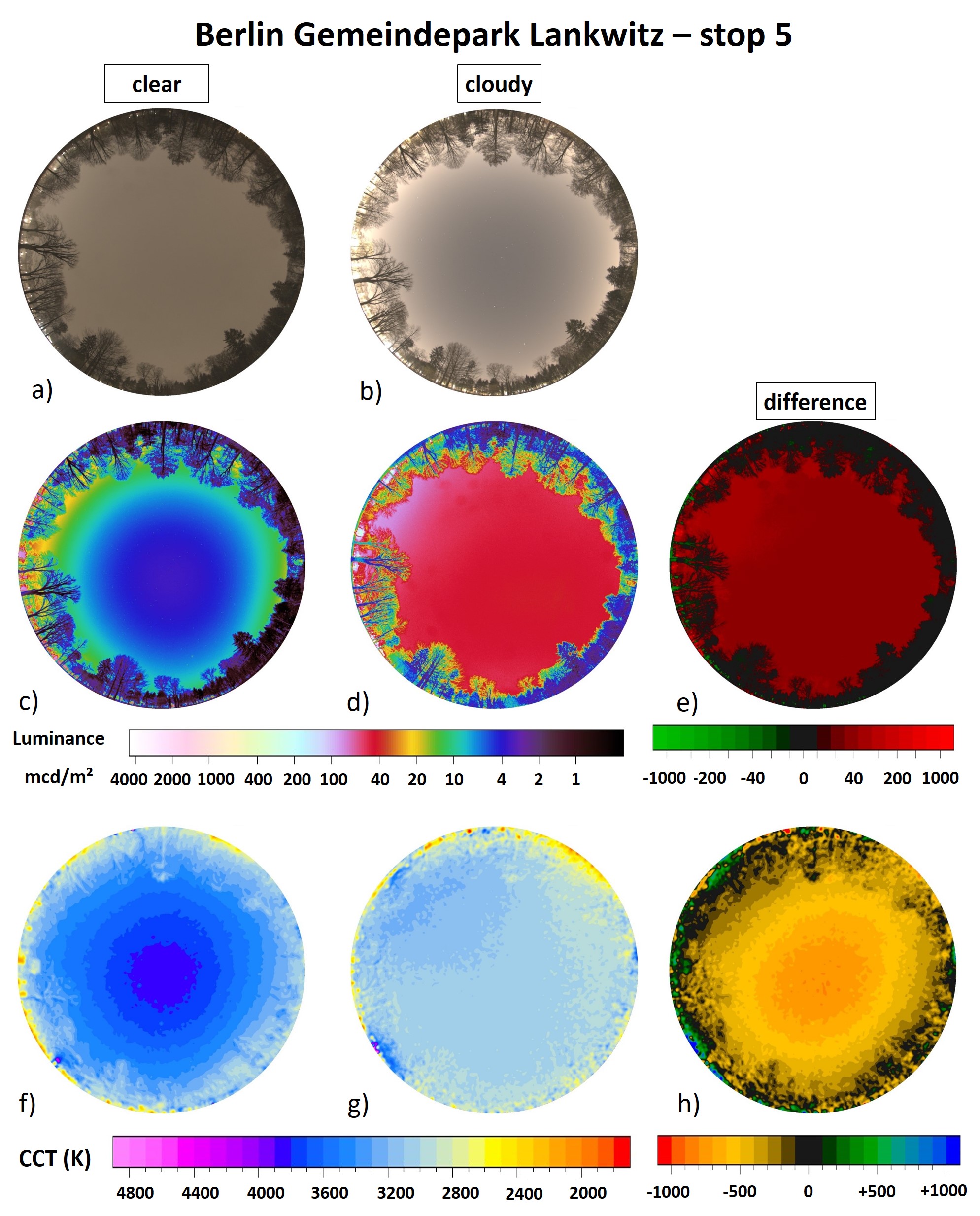}
\caption{All-sky images and calculated imaging data for stop 5 at Berlin, Gemeindepark Lankwitz. The left column shows clear sky data, the middle column overcast sky data and the right hand column subtracted data. a) and b) RGB images; c,d,e luminance; f,g,h CCT.}
\label{stop5}
\end{figure}

\begin{figure}[htp]
\centering
\includegraphics[width=\columnwidth]{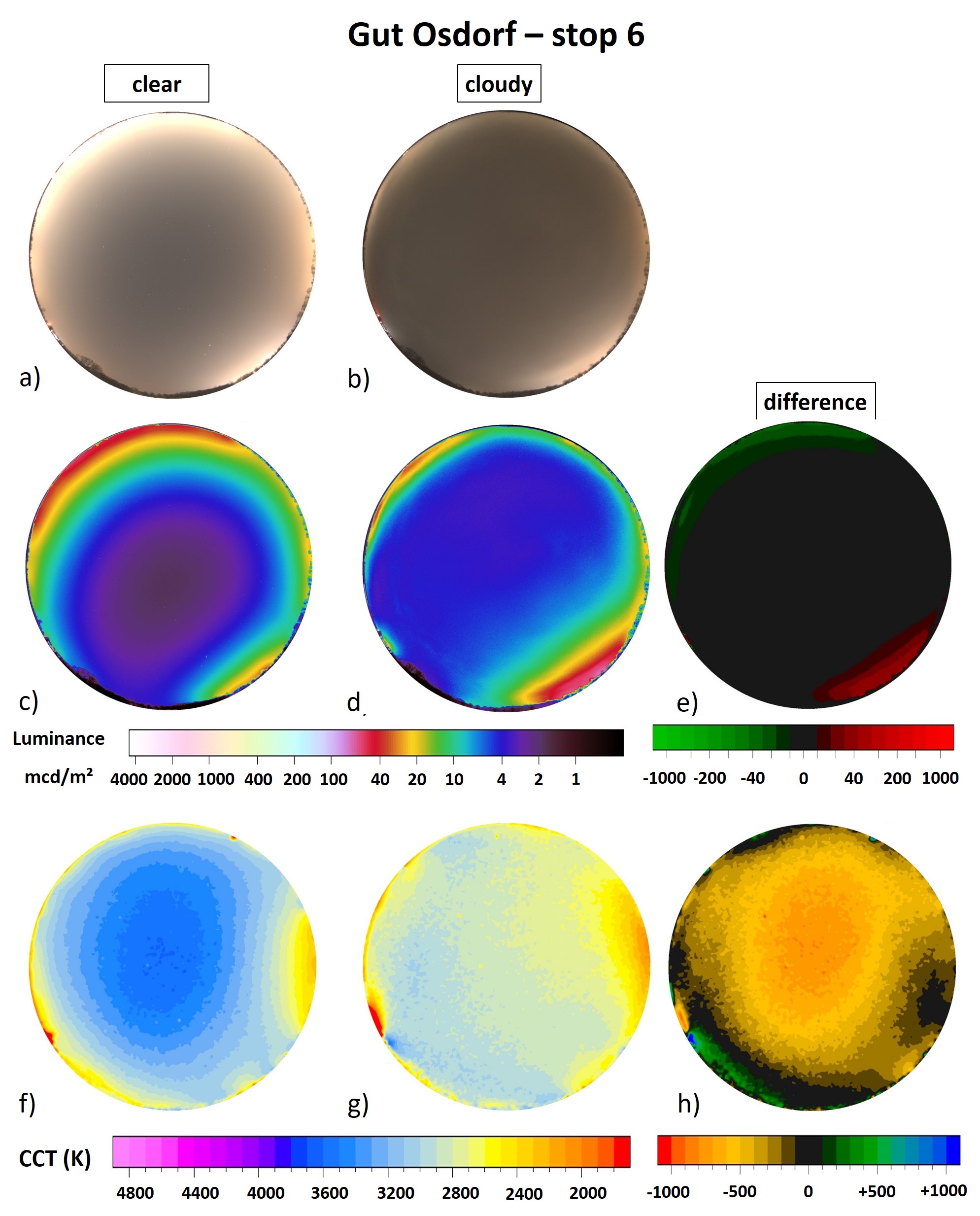}
\caption{All-sky images and calculated imaging data for stop 6 just outside of the Berlin city limit at Gut Osdorf. The left column shows clear sky data, the middle column overcast sky data and the right hand column subtracted data. a) and b) RGB images; c,d,e luminance; f,g,h CCT.}
\label{stop6}
\end{figure}

\begin{figure}[htp]
\centering
\includegraphics[width=\columnwidth]{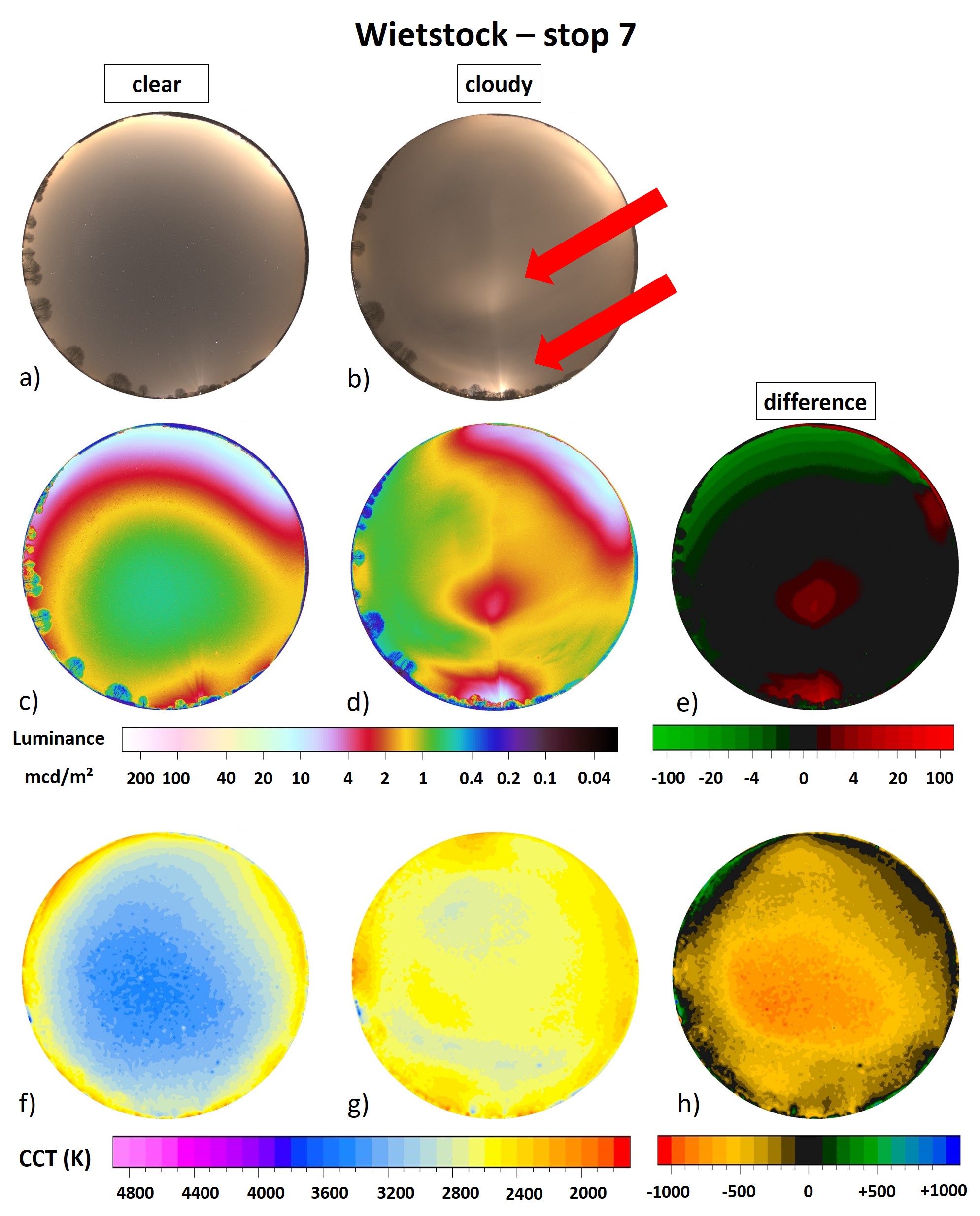}
\caption{All-sky images and calculated imaging data for stop 6 in Wietstock. The left column shows clear sky data, the middle column overcast sky data and the right hand column subtracted data. a) and b) RGB images; c,d,e luminance; f,g,h CCT.}
\label{stop7}
\end{figure}

\begin{figure}[htp]
\centering
\includegraphics[width=\columnwidth]{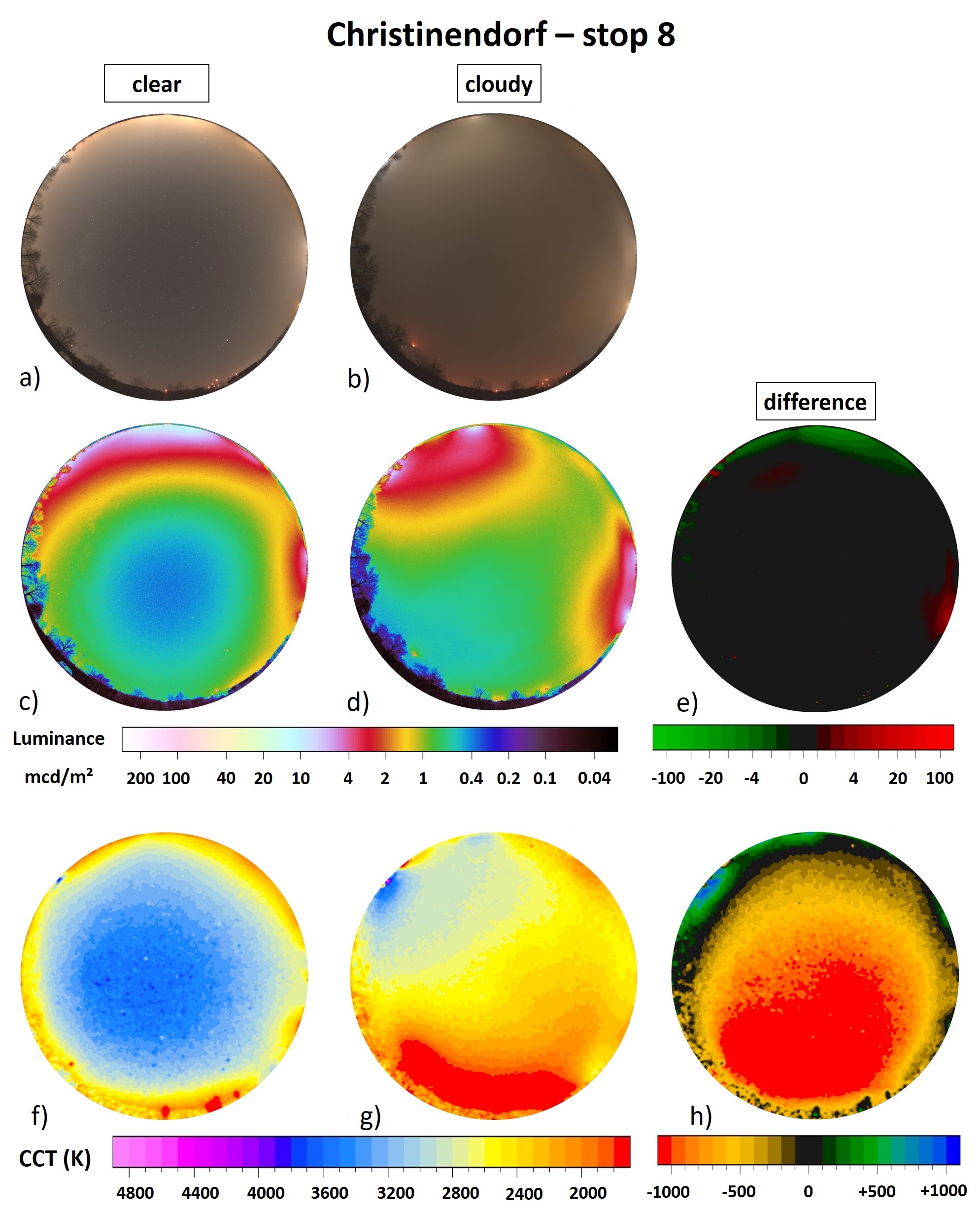}
\caption{All-sky images and calculated imaging data for stop 8 at Christinendorf. The left column shows clear sky data, the middle column overcast sky data and the right hand column subtracted data. a) and b) RGB images; c,d,e luminance; f,g,h CCT. Please note the different luminance scale for the dark sites compared to Fig. 2 and A9-A14.}
\label{stop8}
\end{figure}

\begin{figure}[htp]
\centering
\includegraphics[width=\columnwidth]{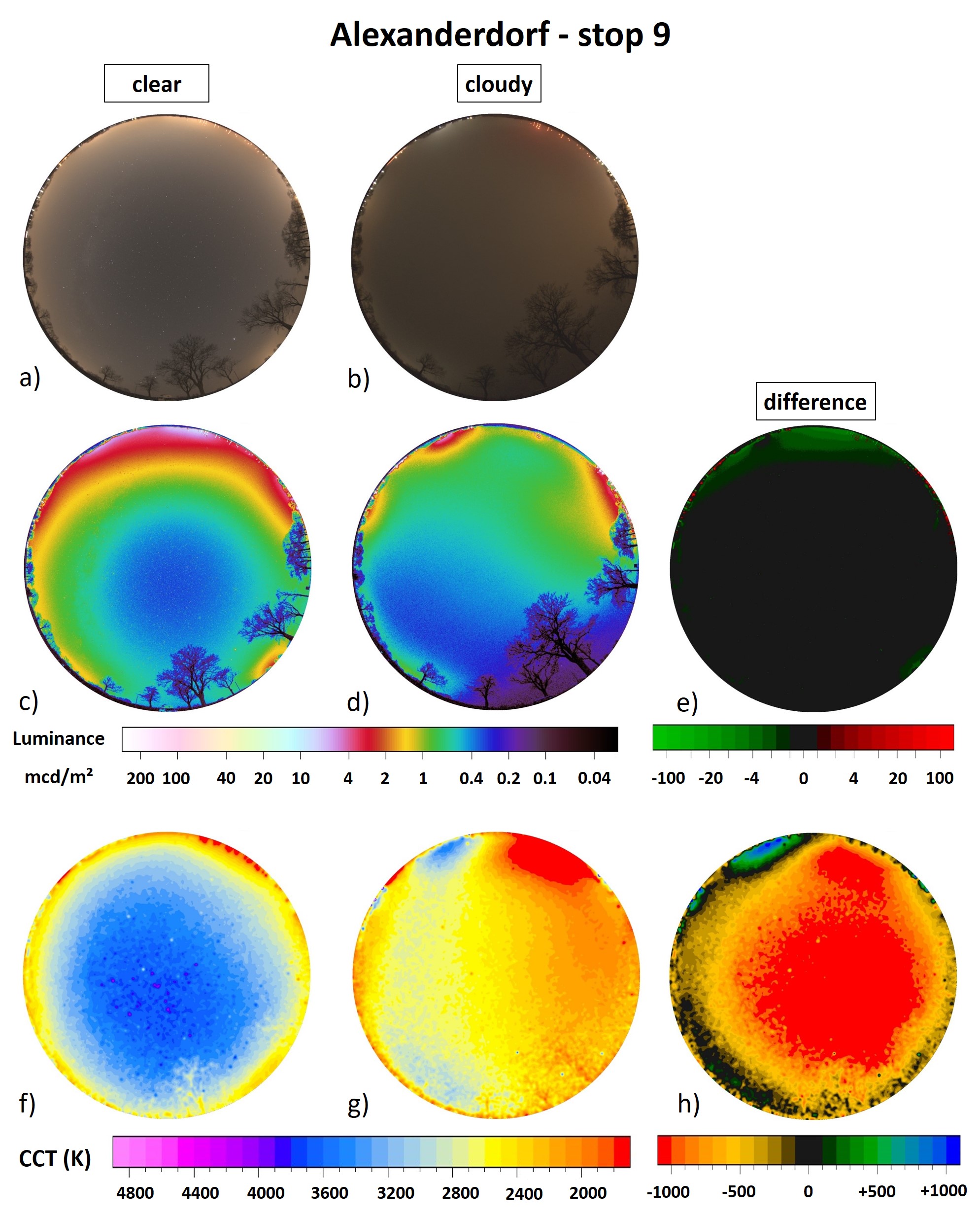}
\caption{All-sky images and calculated imaging data for stop 9 at Alexanderdorf. The left column shows clear sky data, the middle column overcast sky data and the right hand column subtracted data. a) and b) RGB images; c,d,e luminance; f,g,h CCT. Please note the different luminance scale for the dark sites compared to Fig. 2 and A9-A14.}
\label{stop9}
\end{figure}

\begin{figure}[htp]
\centering
\includegraphics[width=\columnwidth]{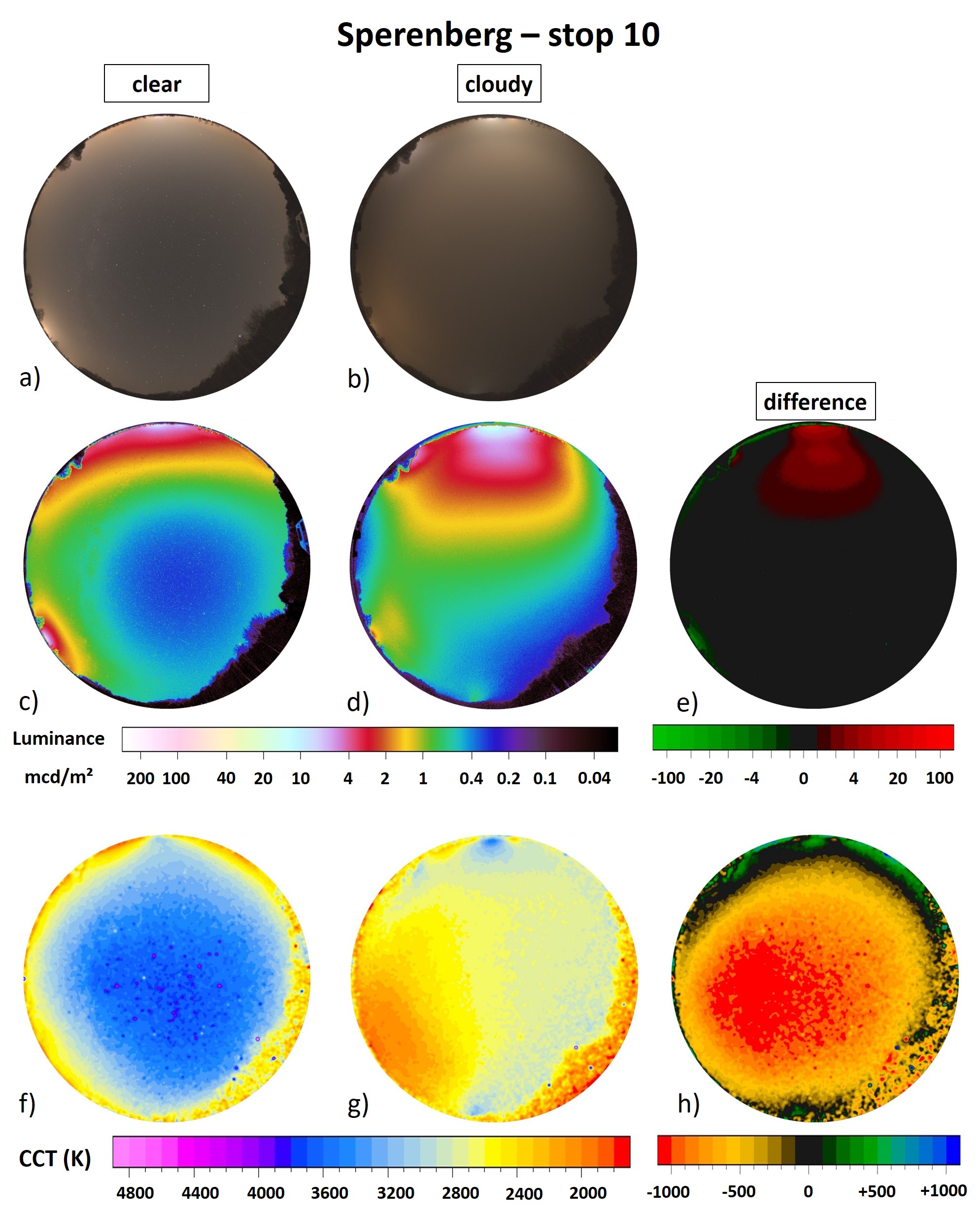}
\caption{All-sky images and calculated imaging data for stop 10 at Sperenberg. The left column shows clear sky data, the middle column overcast sky data and the right hand column subtracted data. a) and b) RGB images; c,d,e luminance; f,g,h CCT. Please note the different luminance scale for the dark sites compared to Fig. 2 and A9-A14.}
\label{stop10}
\end{figure}

\begin{figure}[htp]
\centering
\includegraphics[width=\columnwidth]{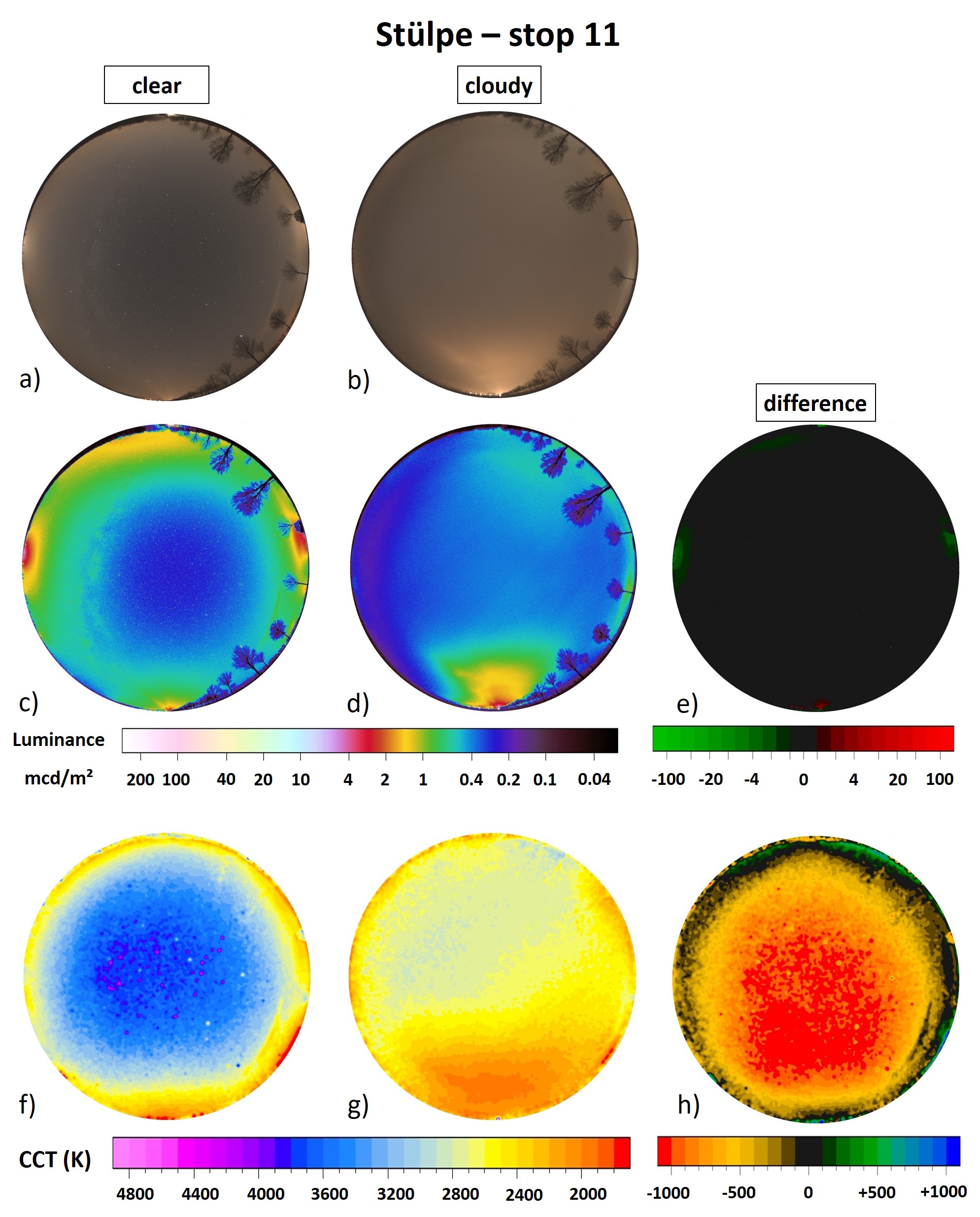}
\caption{All-sky images and calculated imaging data for stop 11 at Stuelpe. The left column shows clear sky data, the middle column overcast sky data and the right hand column subtracted data. a) and b) RGB images; c,d,e luminance; f,g,h CCT. Please note the different luminance scale for the dark sites compared to Fig. 2 and A9-A14.}
\label{stop11}
\end{figure}

\begin{figure}[htp]
\centering
\includegraphics[width=\columnwidth]{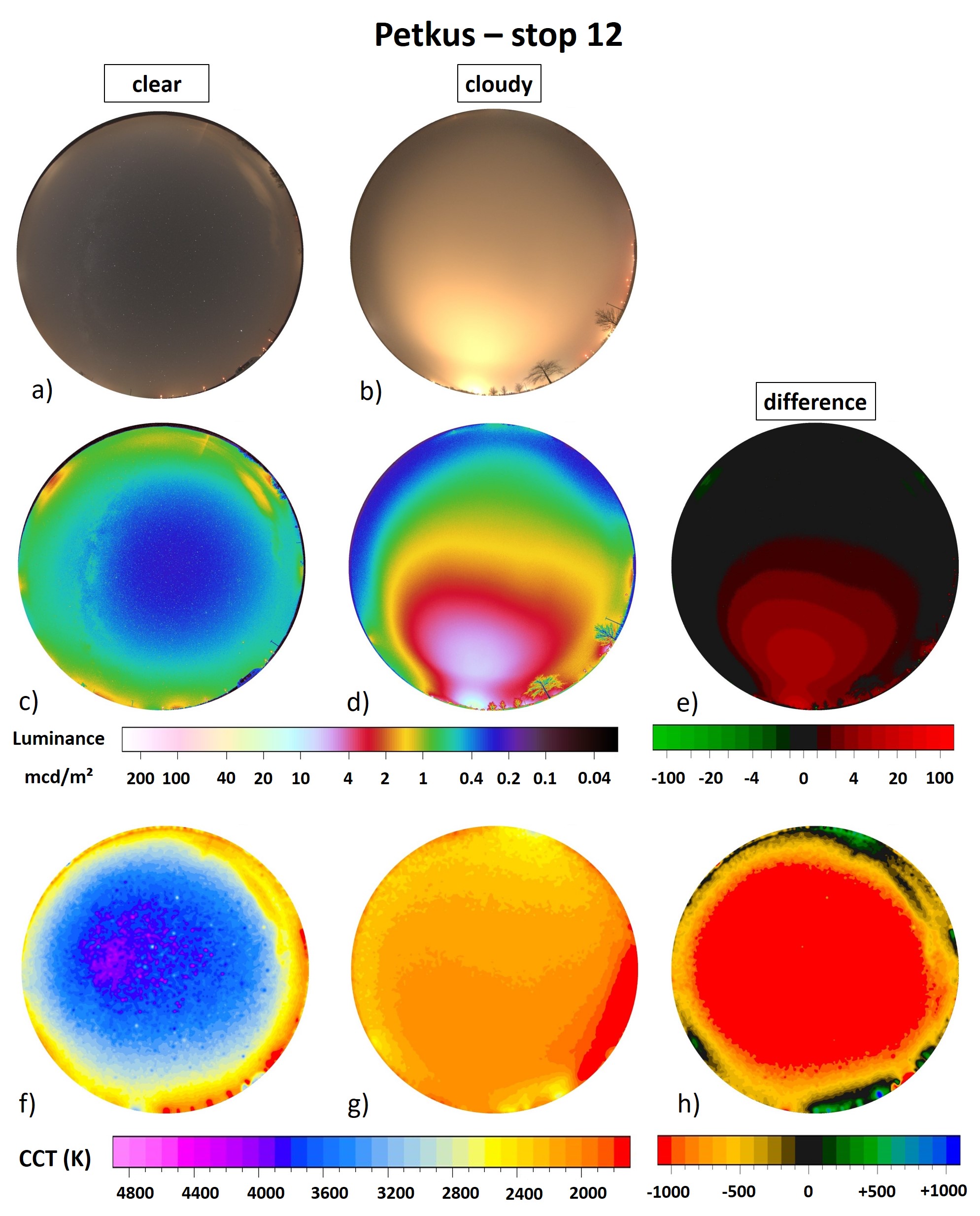}
\caption{All-sky images and calculated imaging data for stop 12 at Petkus. The left column shows clear sky data, the middle column overcast sky data and the right hand column subtracted data. a) and b) RGB images; c,d,e luminance; f,g,h CCT. Please note the different luminance scale for the dark sites compared to Fig. 2 and A9-A14.}
\label{stop12}
\end{figure}

The position and other information of of all stops is listed in table \ref{table_positions} and the full imaging data set from the two transects is shown in figures \ref{stop1}-\ref{stop12}. Each figure is structured as follows: the left column shows data obtained during the clear night, the middle column data obtained during the overcast night and the right hand column data obtained by subtracting the clear night data from the overcast night data. The top row shows RGB images, the middle row luminance data and the lower row CCT data. Please note that the luminance scale is different for the stops within and near the city limit (fig. \ref{stop1}-\ref{stop6} and outside of the city limit (fig. \ref{stop7}-\ref{stop12}).)

\begin{table}[htbp]
  \centering
  \caption{Information about the measurement points.}
  \label{table_positions}
  \begin{tabular}{ccccc}
	&&position & local & time\\
stop Nr.& name & (lat, long) & clear & overcast\\
\hline
- & Alexanderplatz & 52.51942, 13.40647 & - & 1:35 \\
1 & Museumsinsel & 52.52004, 13.39994 & 1:11 & 1:28	\\
2 & Waldeckpark & 52.50654, 13.40359 & 1:38 & 1:50 \\
3 & Park am Gleisdreieck & 52.49472, 13.37880	& 1:57 & 2:03 \\
4 & Hans Baluschek Park & 52.46498, 13.35863	& 2:23 & 2:28 \\
5 & Gemeindepark Lankwitz	& 52.43101, 13.35208 & 2:44 & 2:49 \\
6 & Gut Osdorf &	52.39304, 13.33466	& 3:02 & 3:04 \\
7 & Wietstock & 52.27610, 13.30828 & 3:31 & 3:32\\
8 & Christinendorf & 52.21657, 13.29815	& 3:56 & 4:15 \\
9 & Alexanderdorf & 52.17978, 13.31725	&	4:05 & 4:23	\\
10 & Sperenberg	& 52.12899, 13.36726 & 4:19 & 3:31	\\
11 & St{\"u}lpe & 52.06222, 13.32610 & 4:31 & 4:43 \\
12 & Petkus	& 51.99443, 13.35099	& 4:47 & 4:55\\
  \end{tabular}
\end{table}

\begin{figure}[htp]
\centering
\includegraphics[width=\columnwidth]{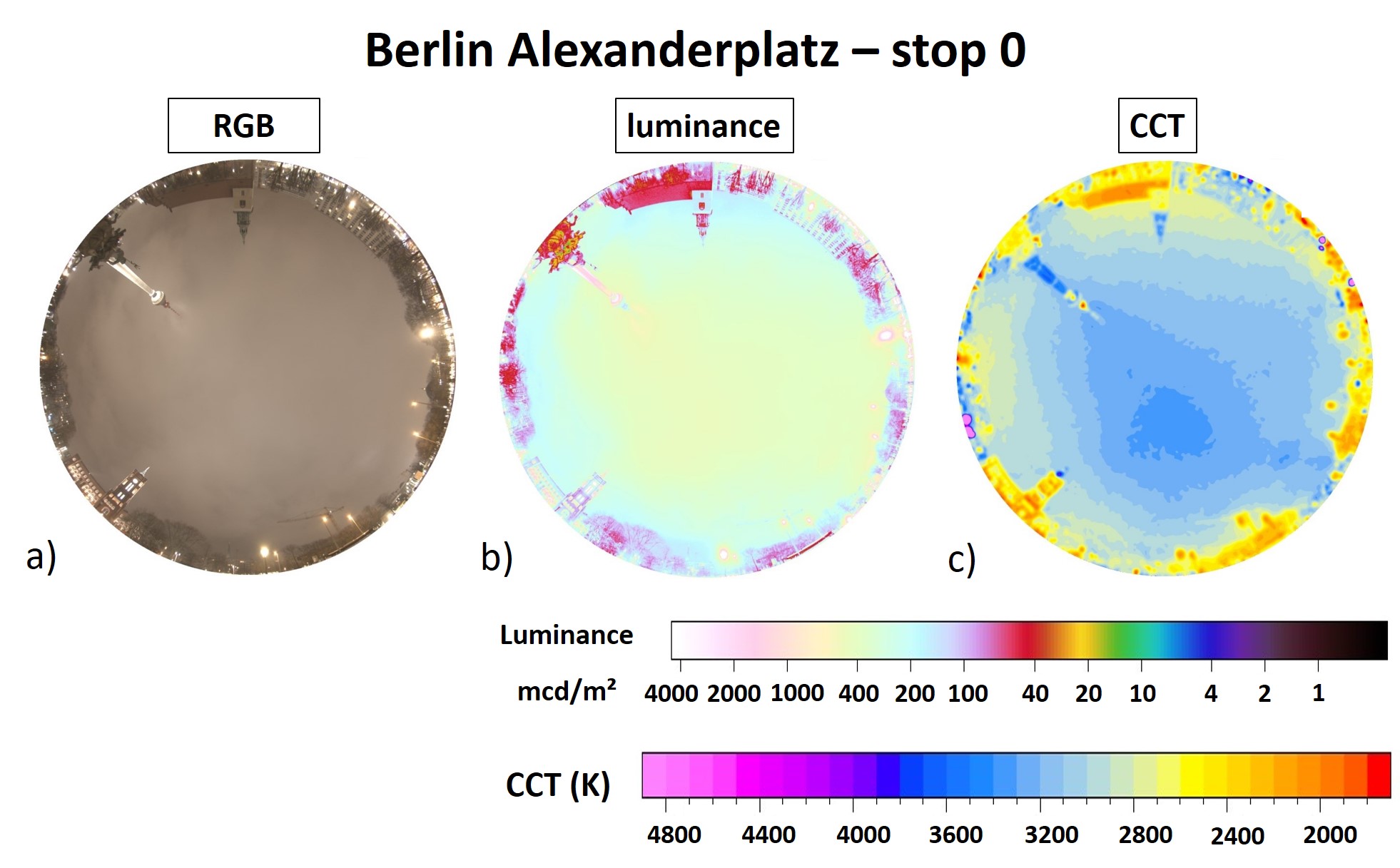}
\caption{All-sky image and calculated imaging data for stop 0 at Berlin Alexanderplatz for overcast conditions. The left column a) shows the RGB image, the middle column b) the luminance map and the right hand column c) CCT map.}
\label{stop0}
\end{figure}

\pagebreak
\section{Time series data}
\begin{figure}[htp]
\centering
a)\includegraphics[width=0.75\columnwidth]{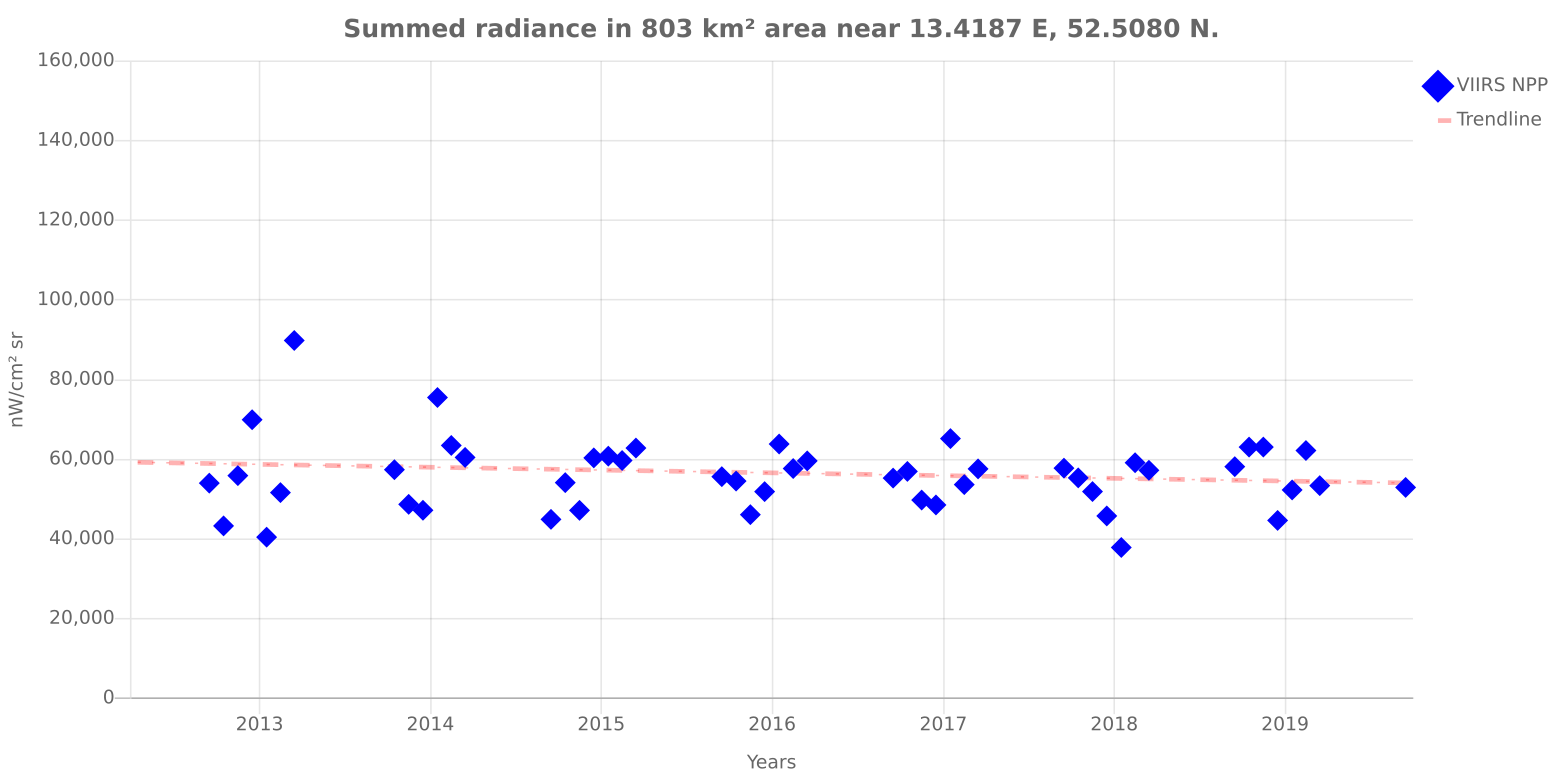}\\
b)\includegraphics[width=0.75\columnwidth]{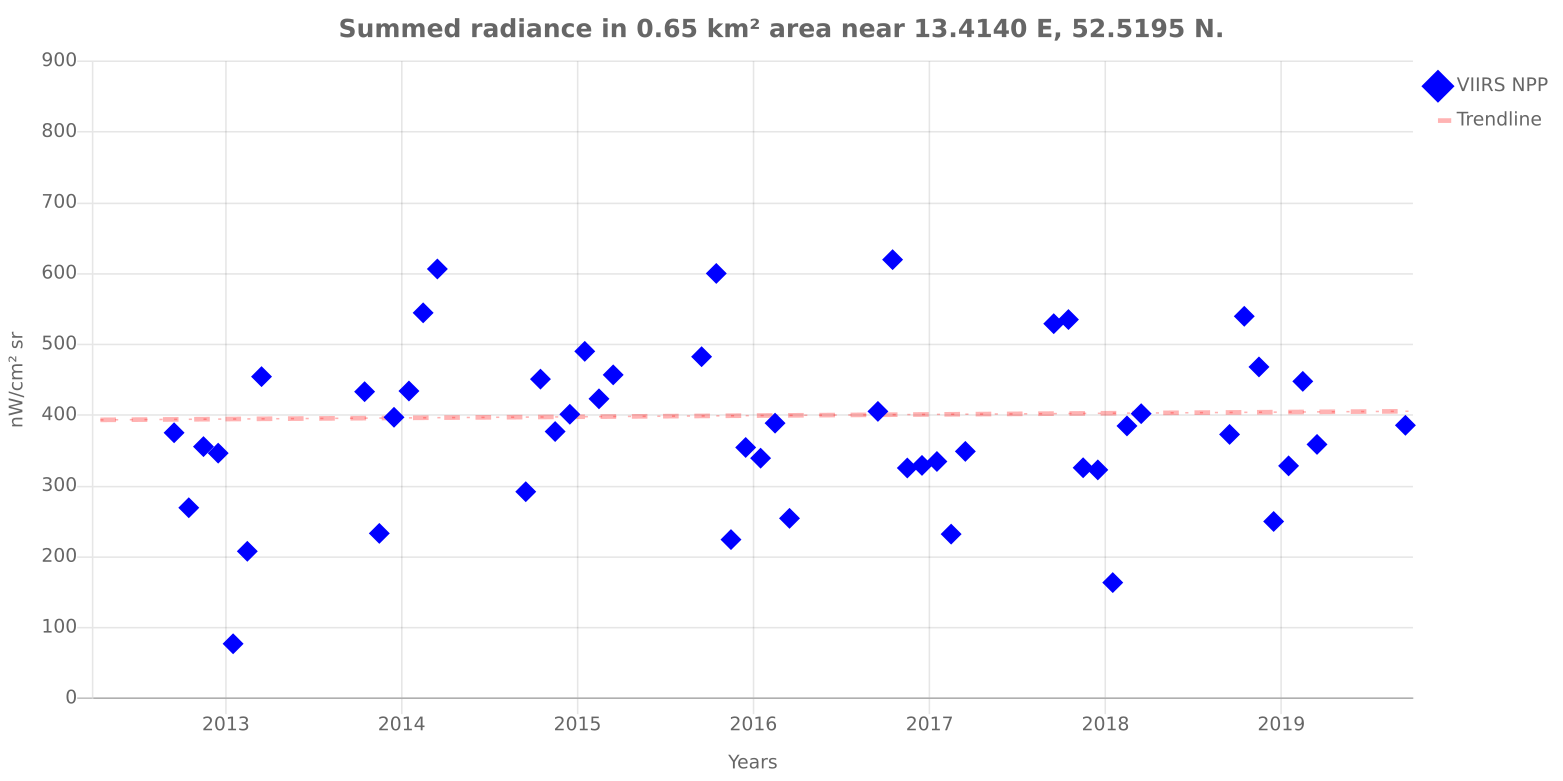}
\caption{Radiance time series showing monthly composites of VIIRS/DNB between 2012 and 2018 a) Berlin and b) Berlin, Alexanderplatz. Data from https://lighttrends.lightpollutionmap.info.}
\label{lighttrends1}
\end{figure}

\begin{figure}[htp]
\centering
a)\includegraphics[width=0.75\columnwidth]{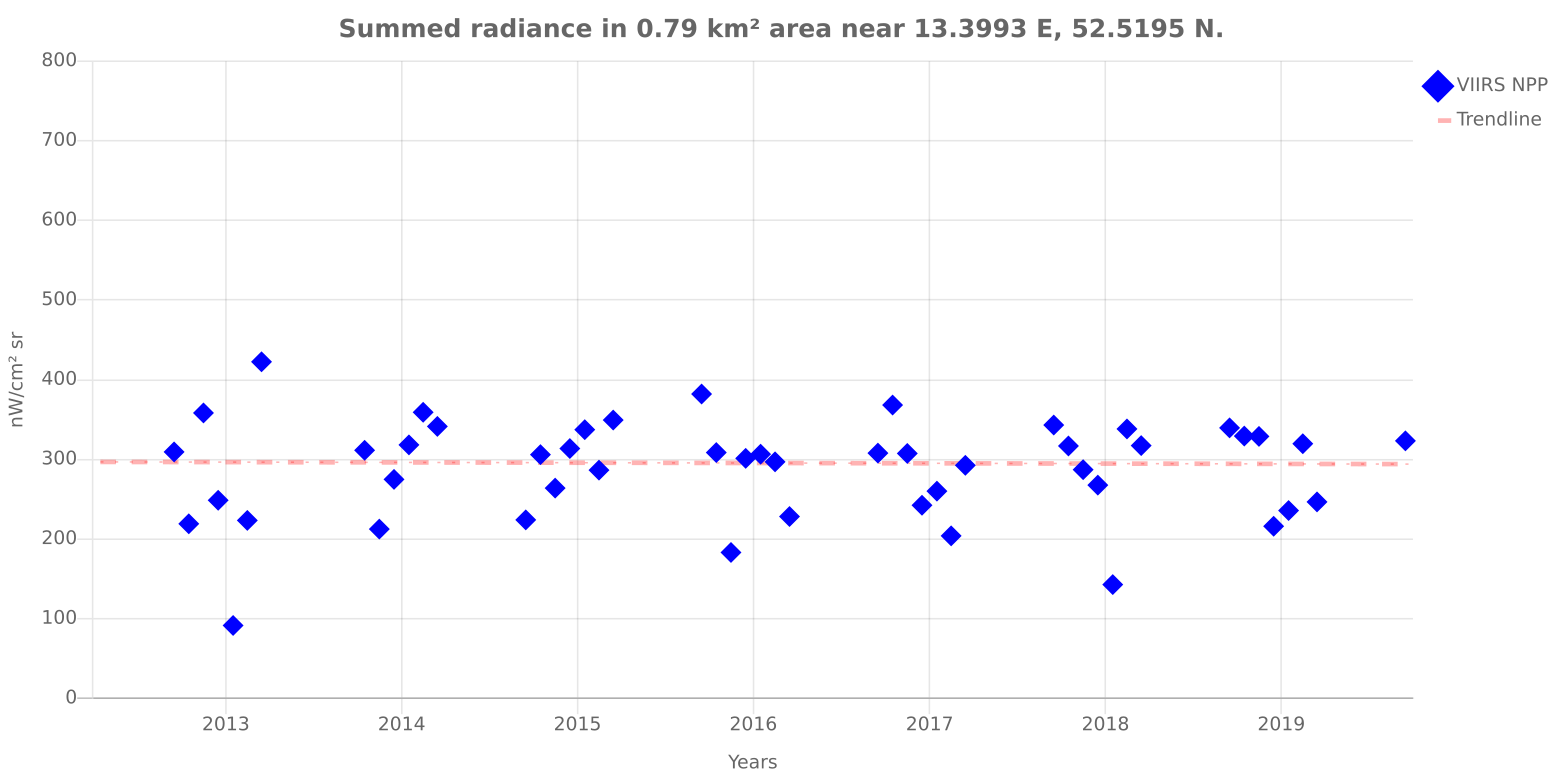}\\
b)\includegraphics[width=0.75\columnwidth]{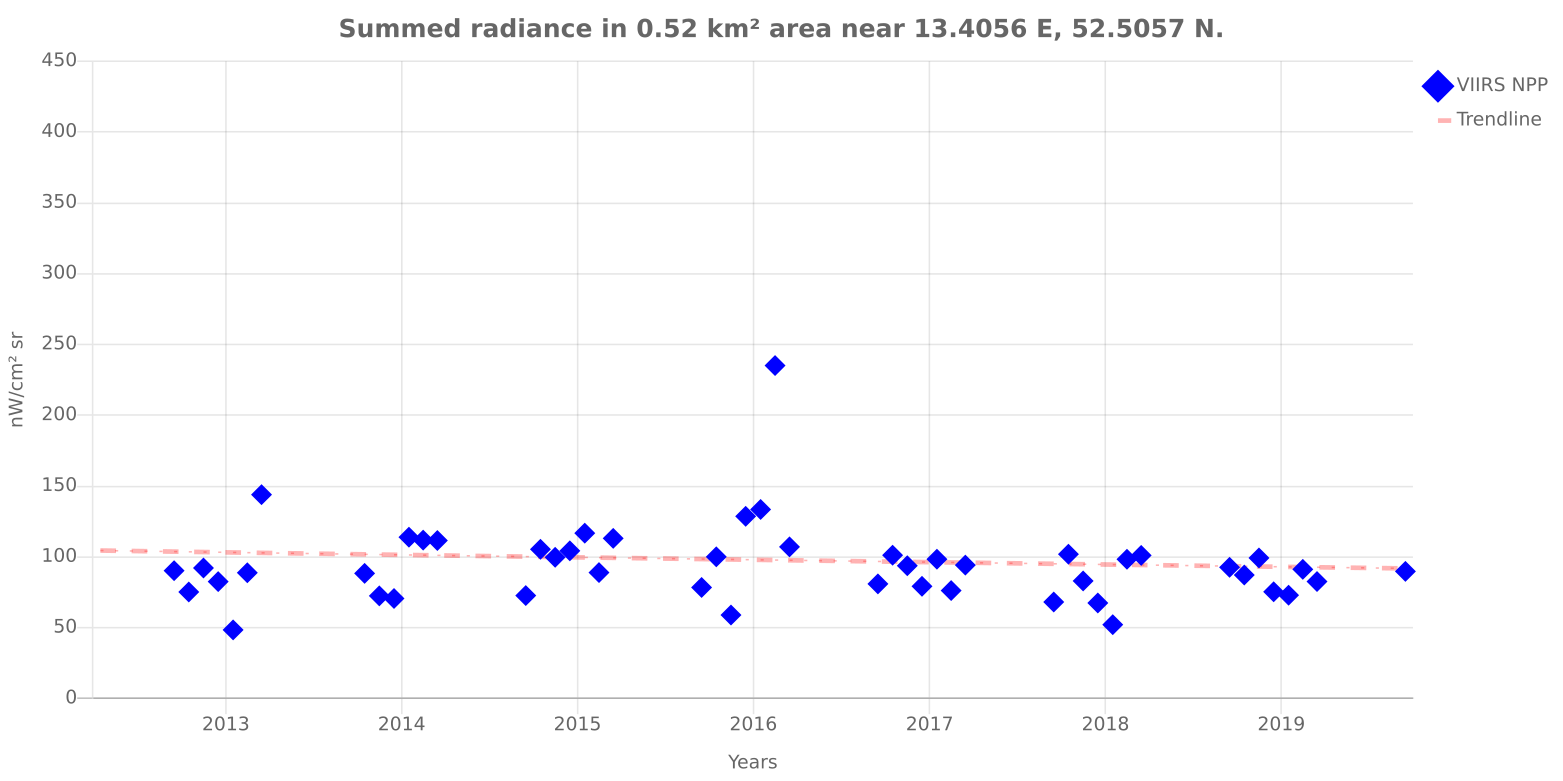}\\
c)\includegraphics[width=0.75\columnwidth]{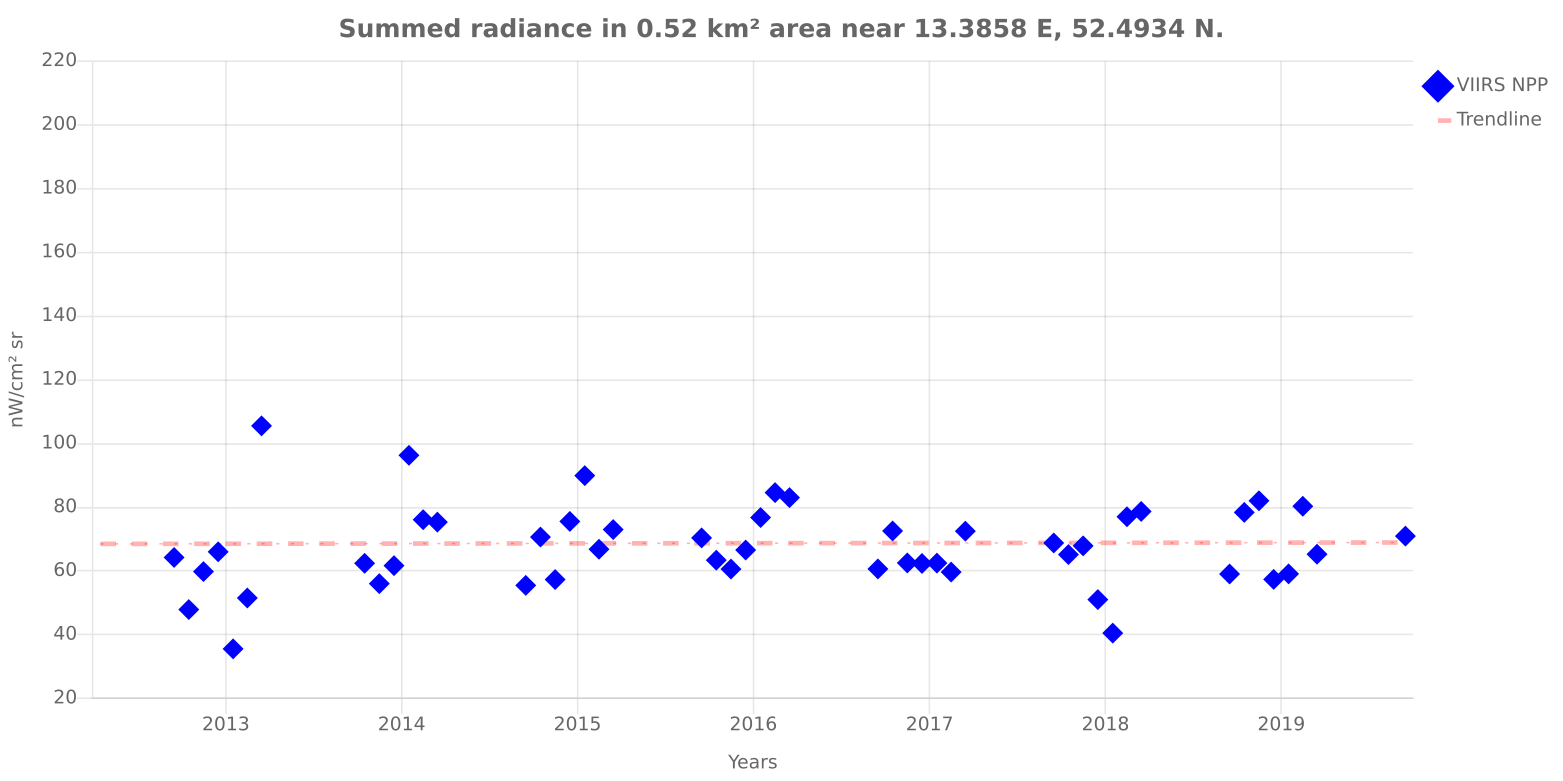}
\caption{Radiance time series showing monthly composites of VIIRS/DNB between 2012 and 2018 for the stops of the transect a) Museumsinsel b) Waldeckpark c) Park am Gleisdreieck. Data from https://lighttrends.lightpollutionmap.info.}
\label{lighttrends2}
\end{figure}

\begin{figure}[htp]
\centering
a)\includegraphics[width=0.75\columnwidth]{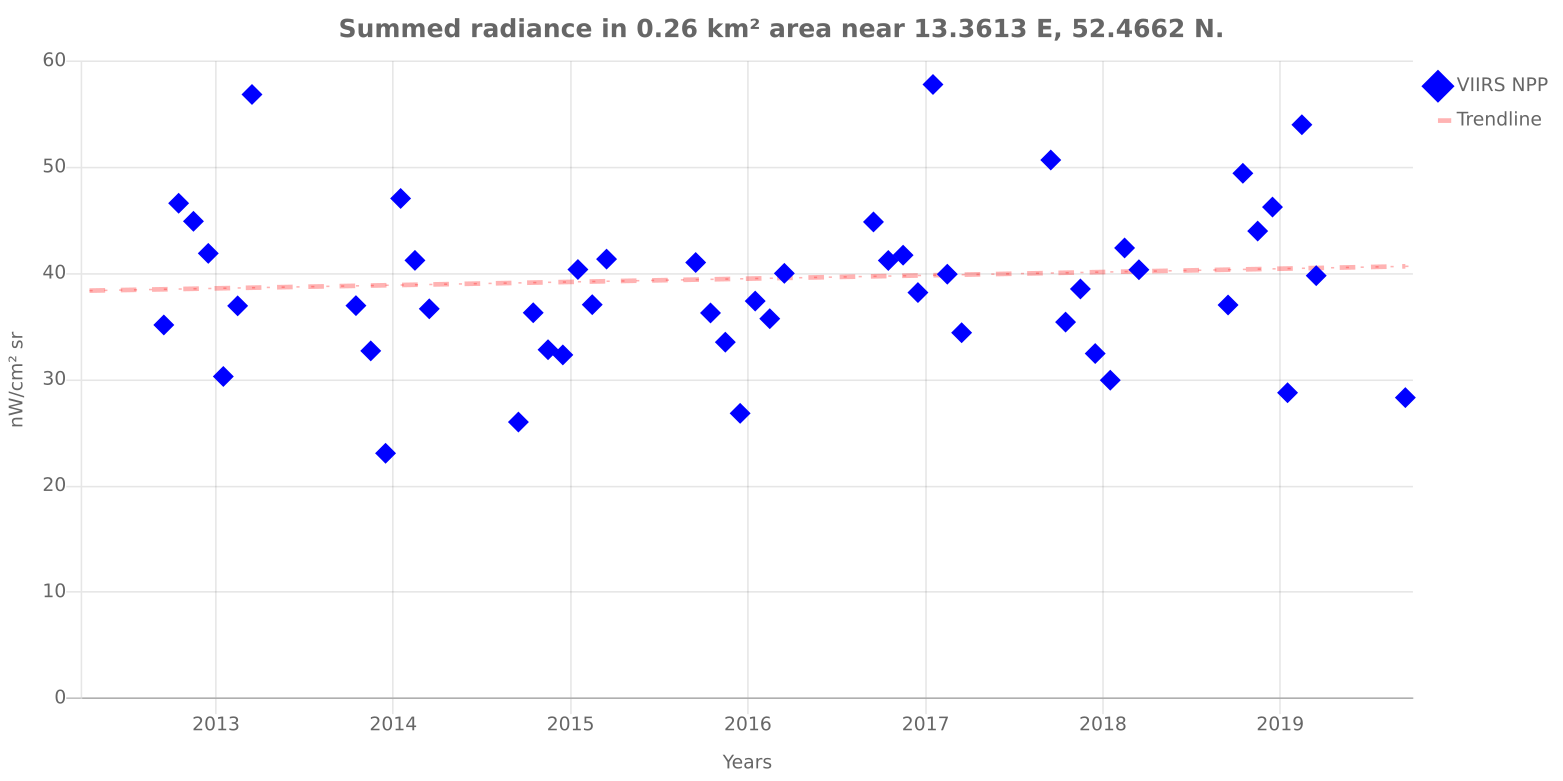}\\
b)\includegraphics[width=0.75\columnwidth]{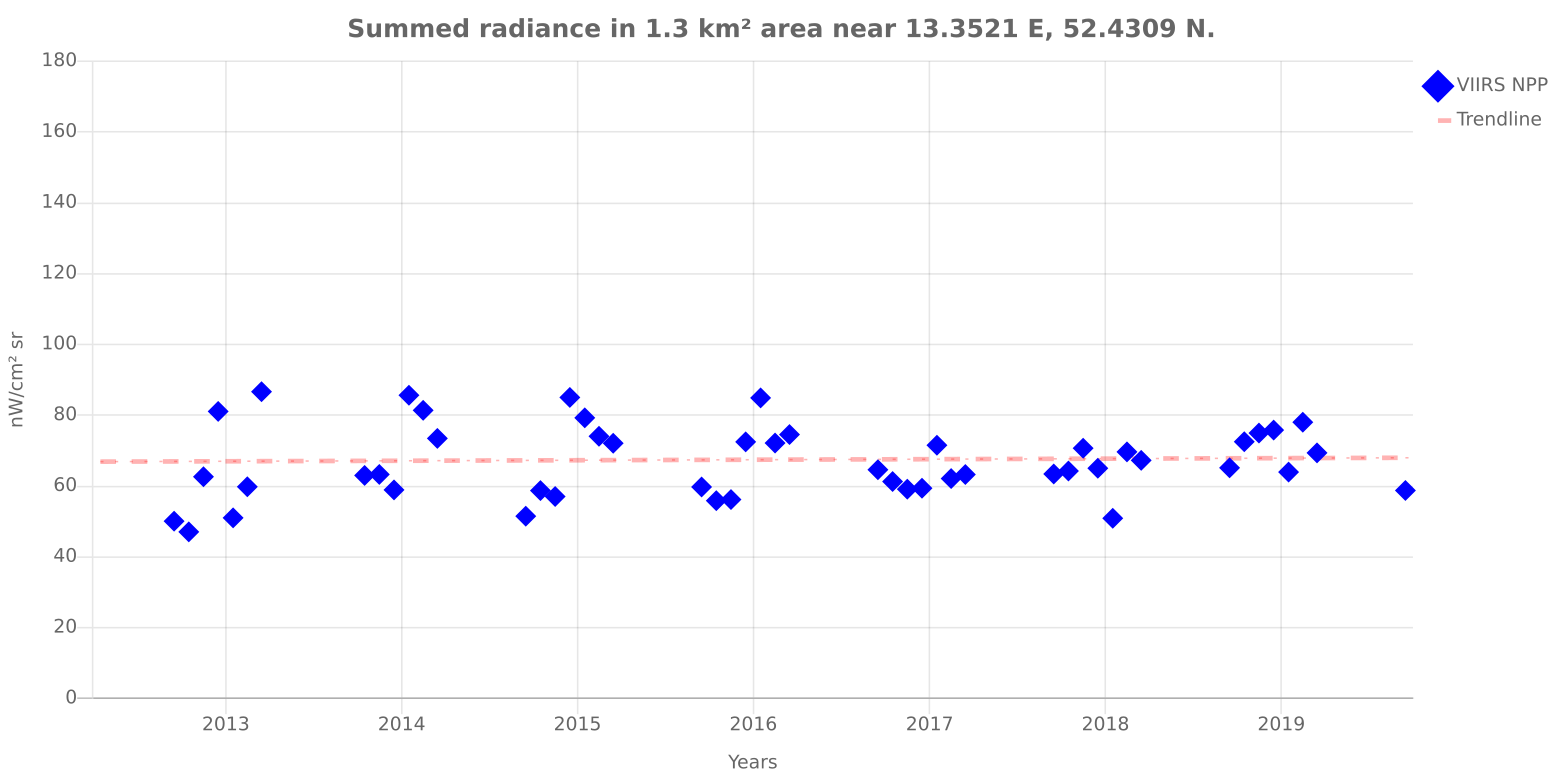}\\
c)\includegraphics[width=0.75\columnwidth]{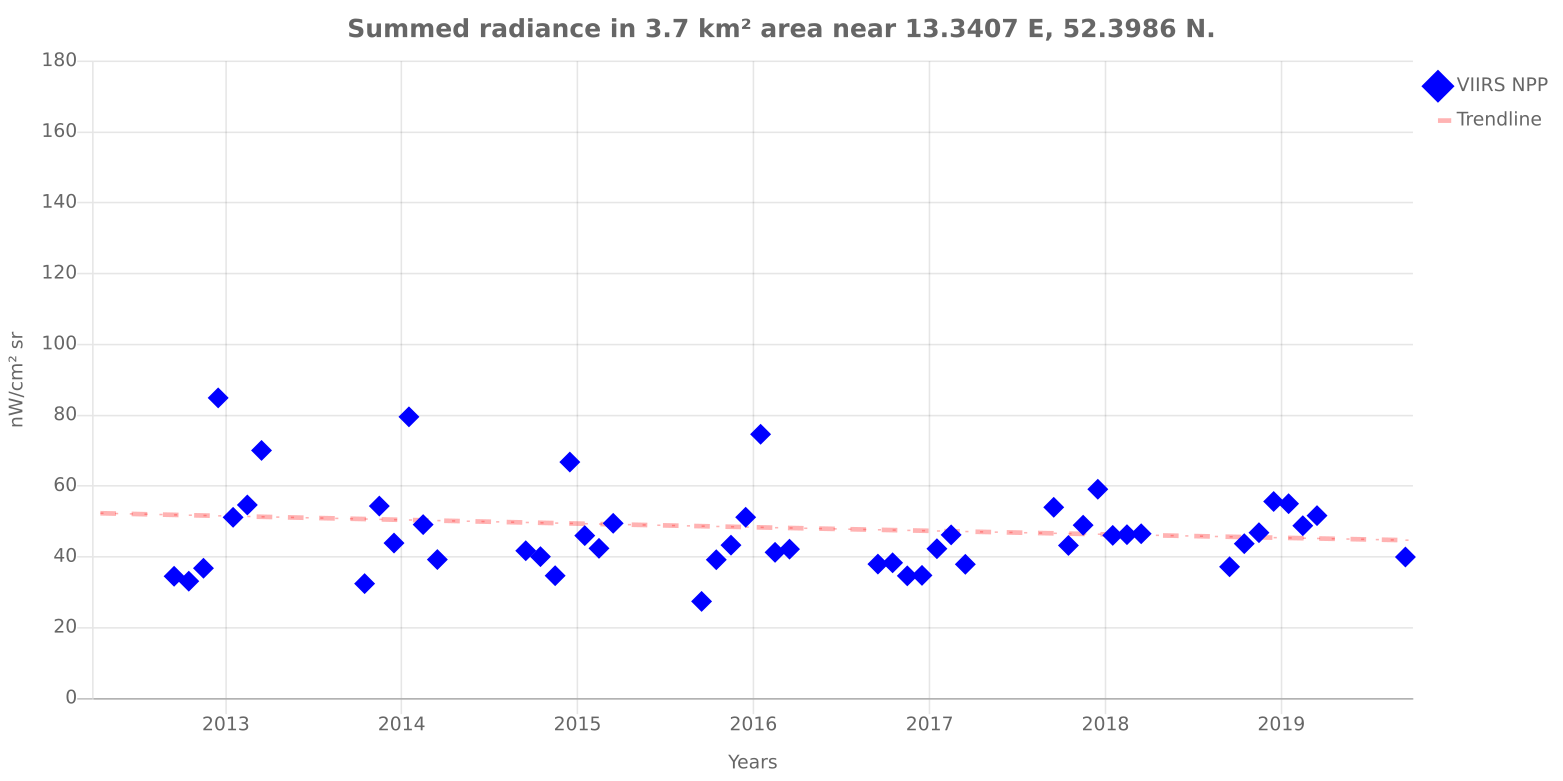}\\
\caption{Radiance time series showing monthly composites of VIIRS/DNB between 2012 and 2018 for the stops of the transect a) Baluschek Park b) Gemeindepark Lankwitz c) Gut Osdorf. Data from https://lighttrends.lightpollutionmap.info.}
\label{lighttrends3}
\end{figure}

\begin{figure}[htp]
\centering
a)\includegraphics[width=0.75\columnwidth]{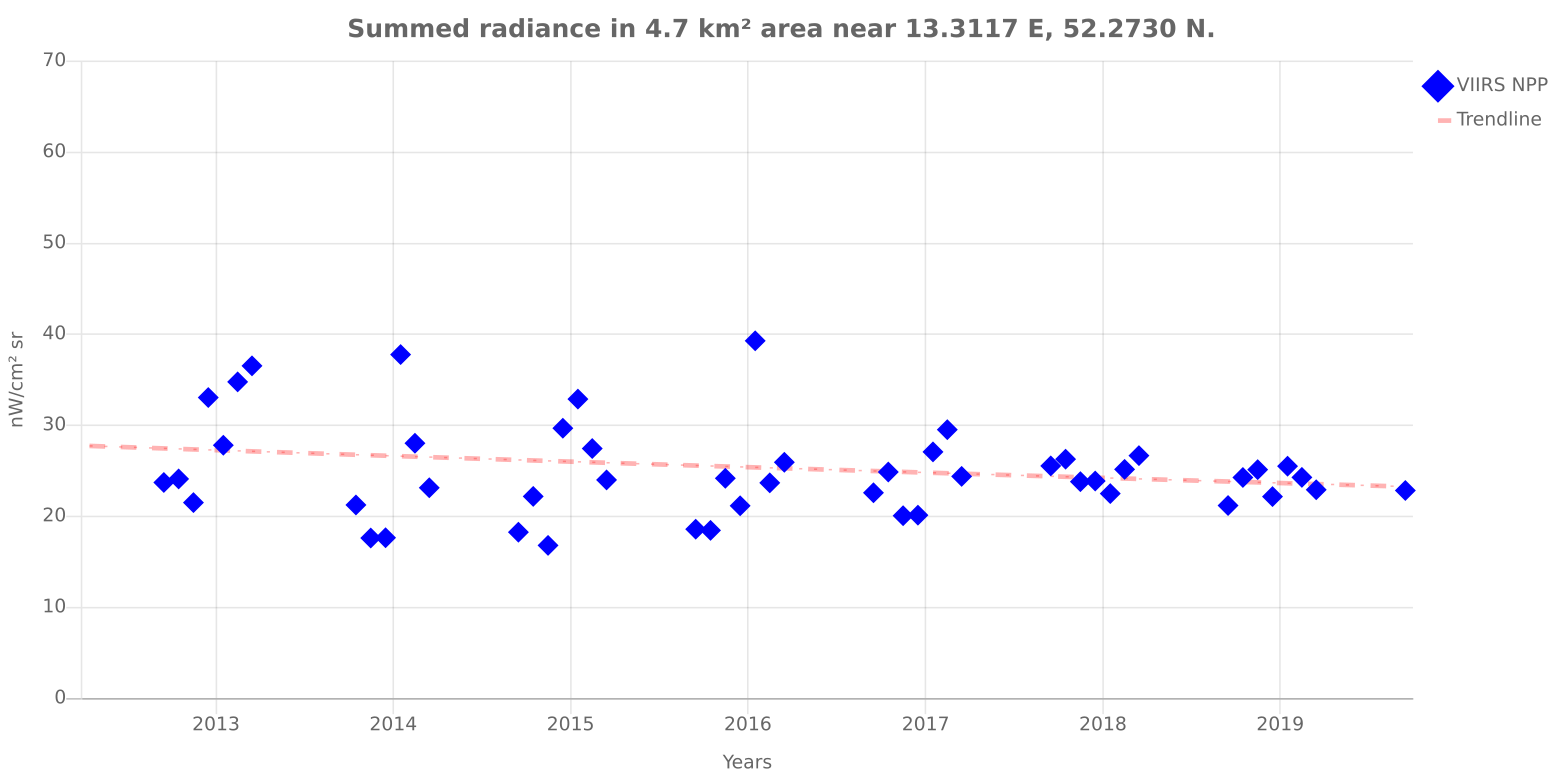}\\
b)\includegraphics[width=0.75\columnwidth]{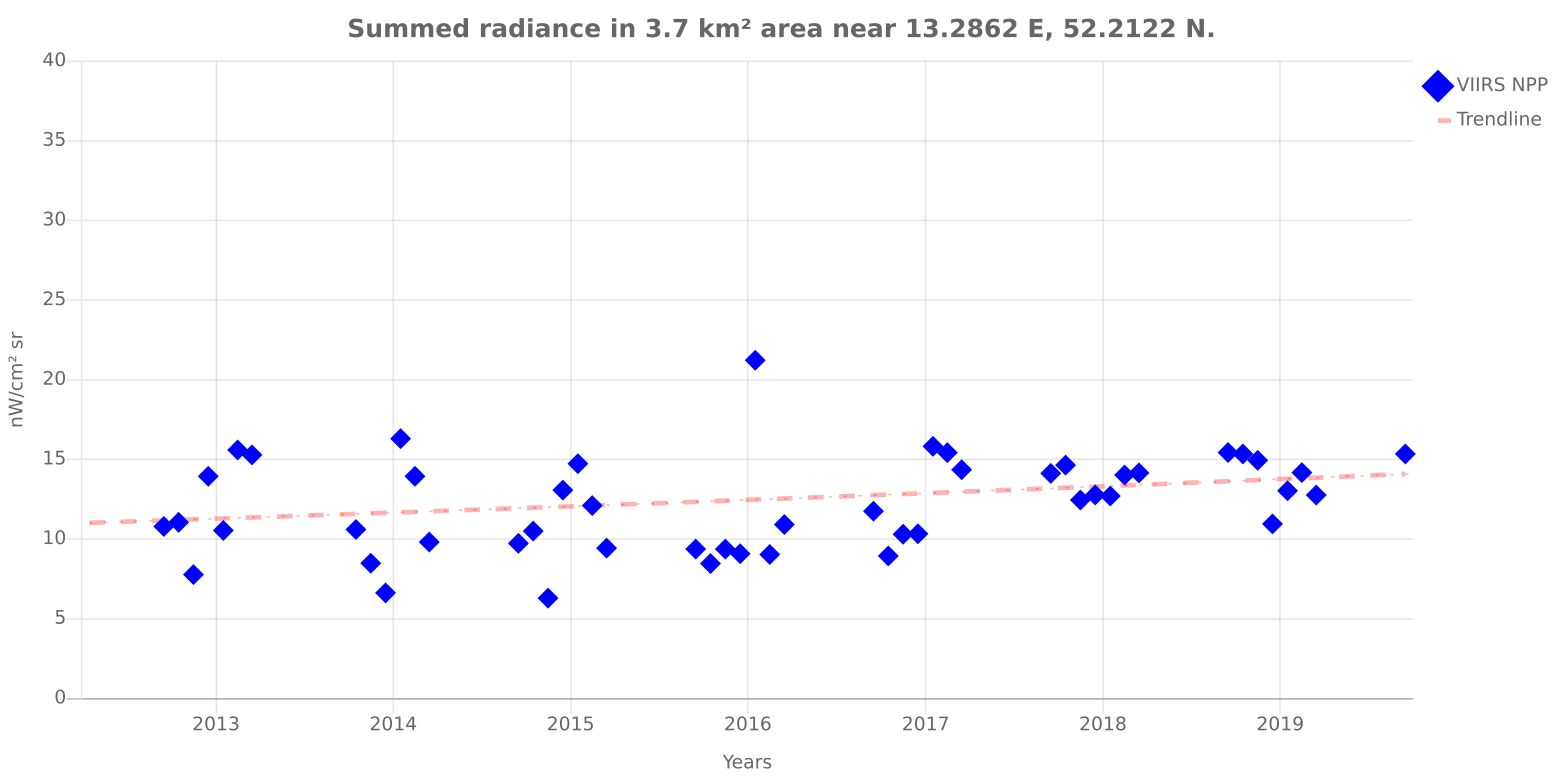}\\
c)\includegraphics[width=0.75\columnwidth]{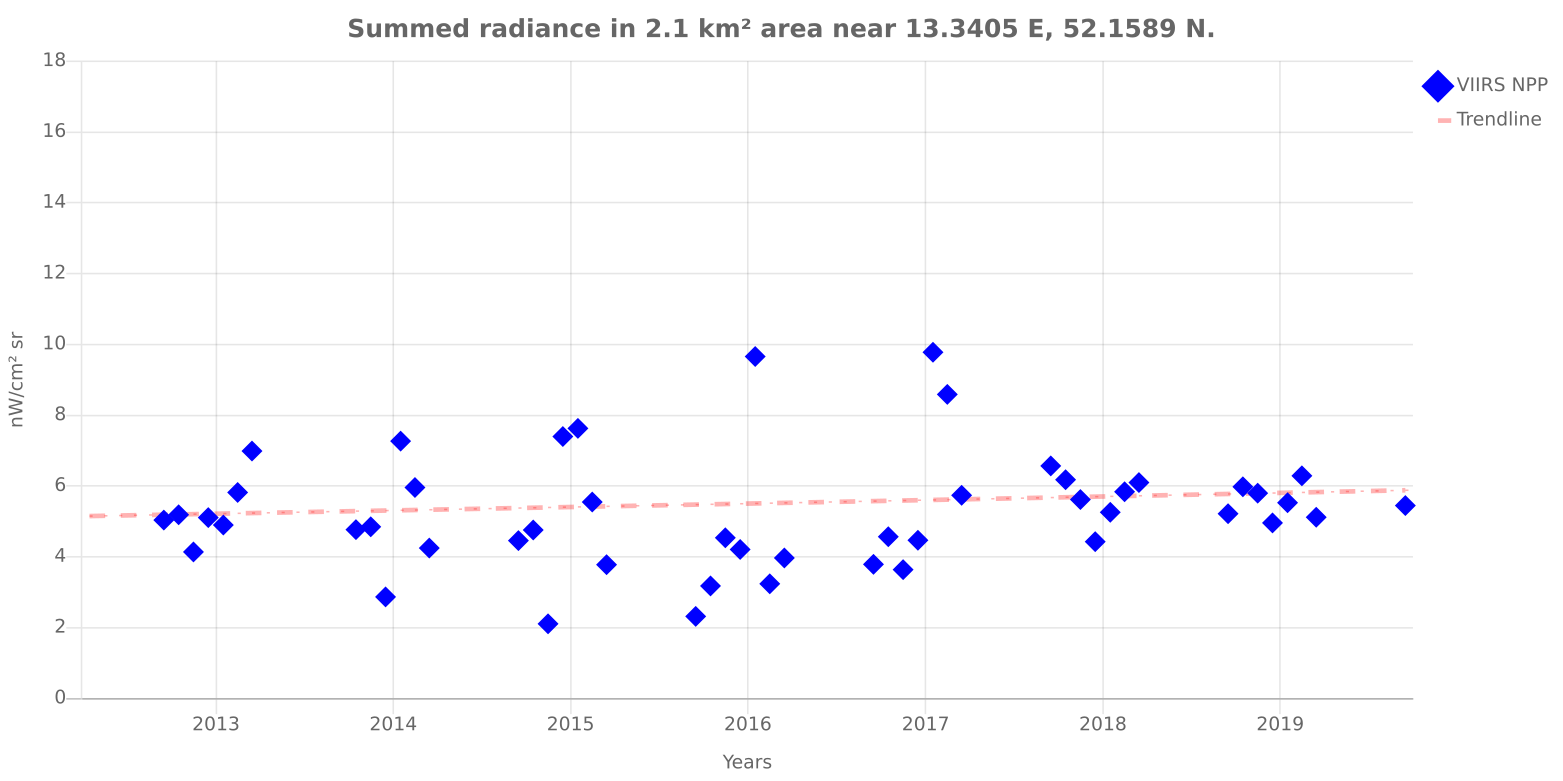}
\caption{Radiance time series showing monthly composites of VIIRS/DNB between 2012 and 2018 for the stops of the transect a) Wietstock b) Christinendorf c) Alexanderdorf. Data from https://lighttrends.lightpollutionmap.info.}
\label{lighttrends4}
\end{figure}

\begin{figure}[htp]
\centering
a)\includegraphics[width=0.75\columnwidth]{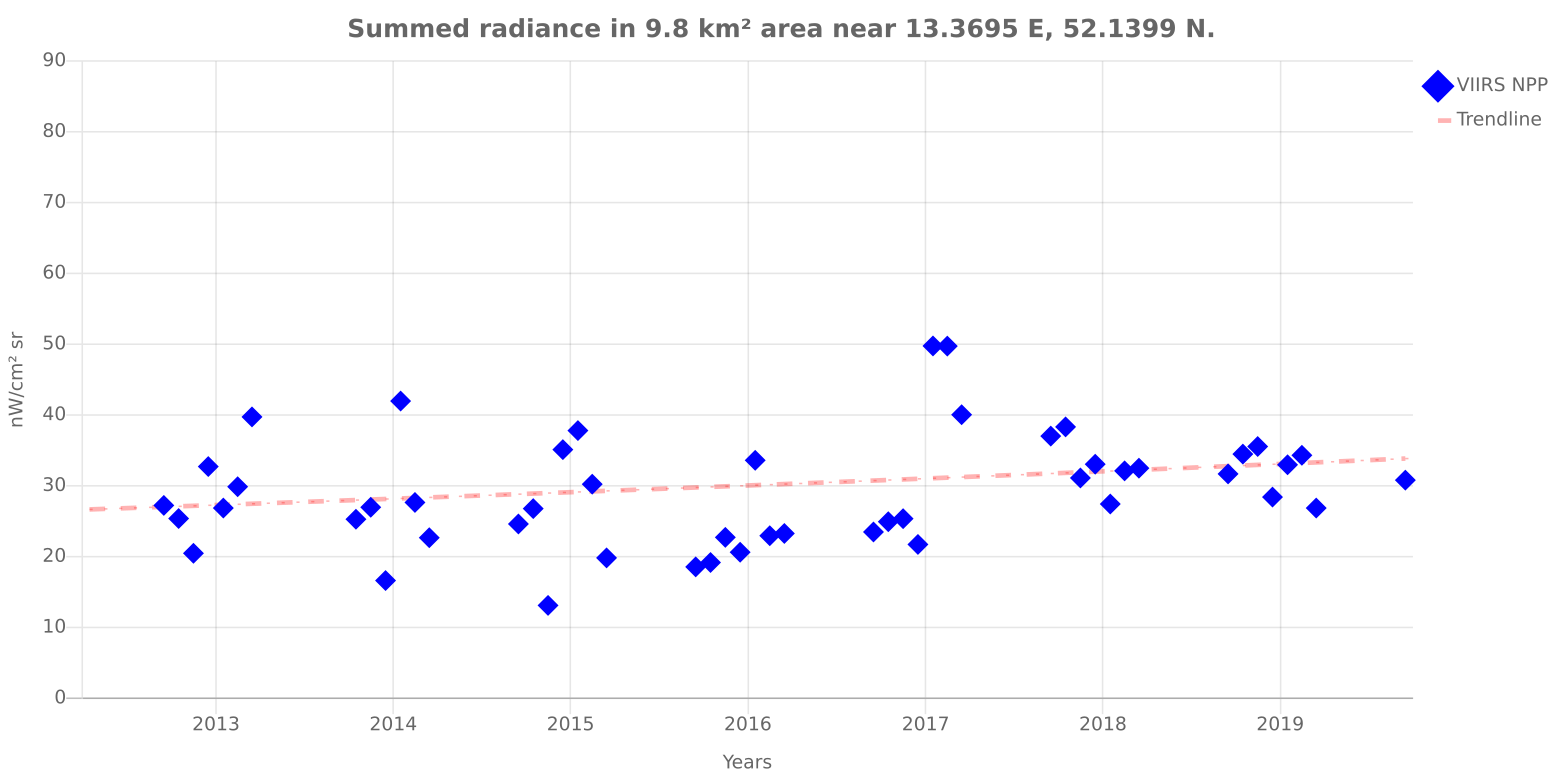}\\
b)\includegraphics[width=0.75\columnwidth]{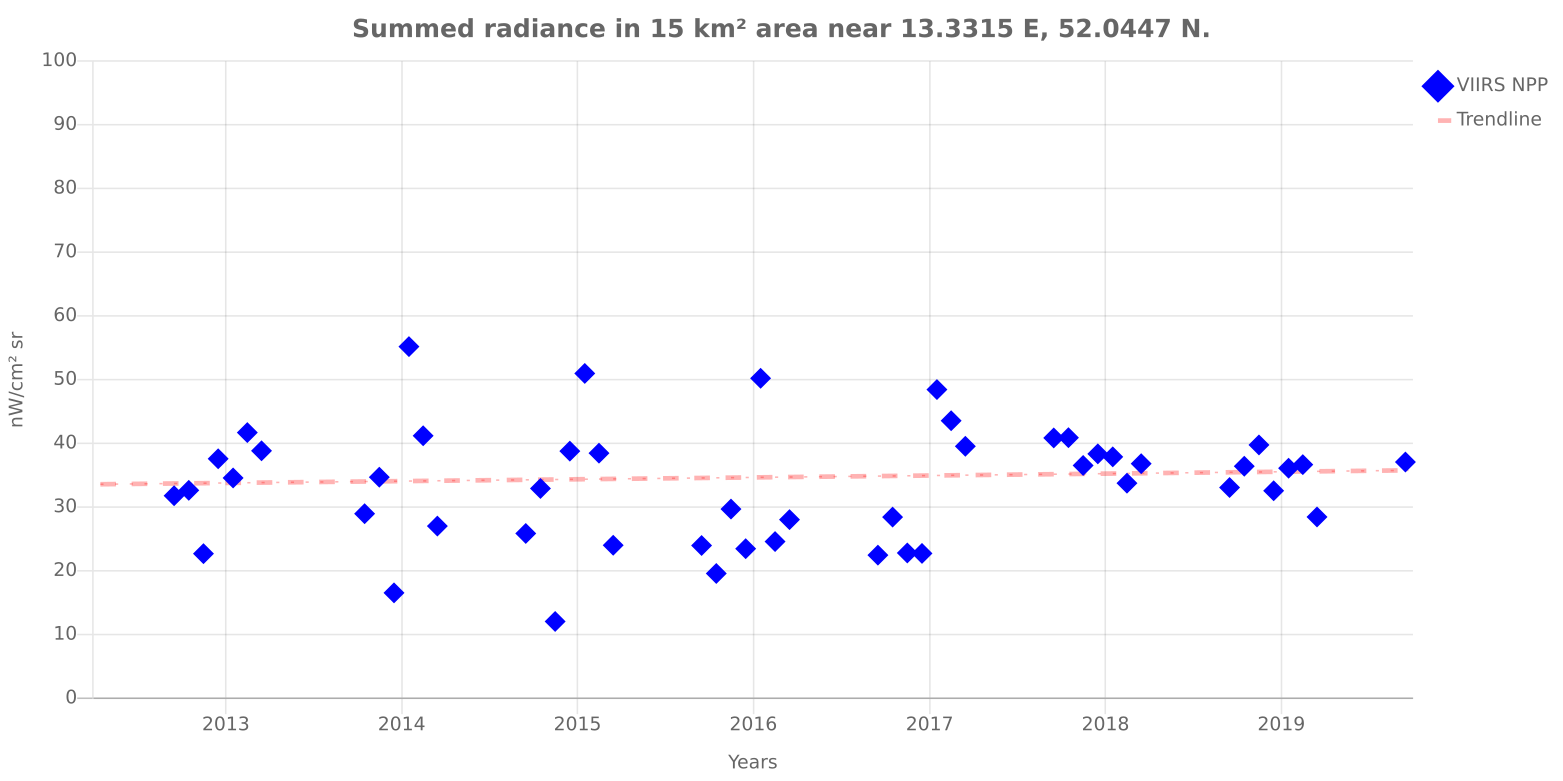}\\
c)\includegraphics[width=0.75\columnwidth]{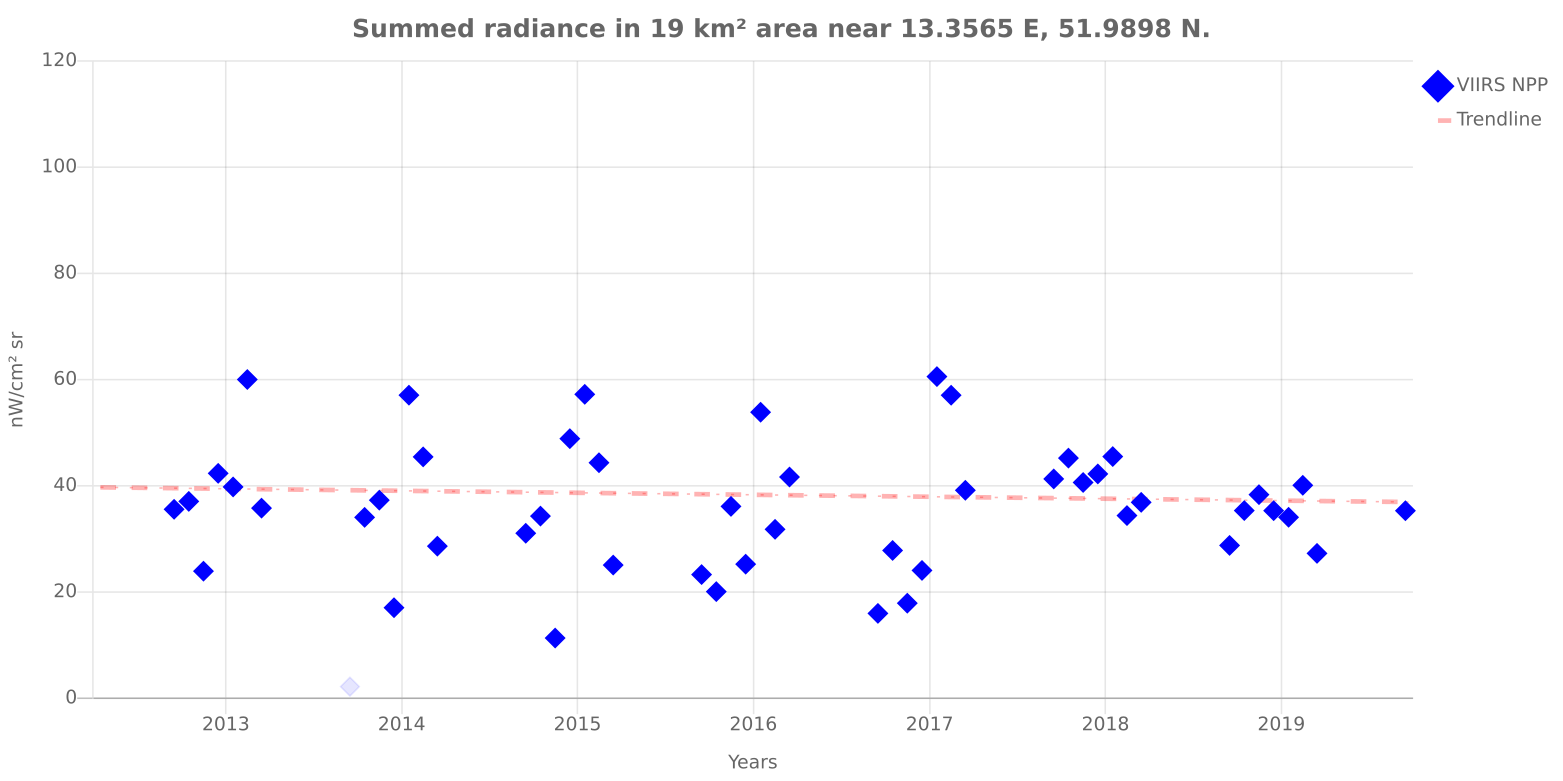}\\
\caption{Radiance time series showing monthly composites of VIIRS/DNB between 2012 and 2018 for the stops of the transect a) Sperenberg b) St{\"u}lpe c) Petkus. Data from https://lighttrends.lightpollutionmap.info.}
\label{lighttrends5}
\end{figure}

Time series data of the upwelling night-time radiance detected by VIIRS/DNB using monthly composites (not all months of the year are used, see Hyde and Coesfeld for discussions) are shown in fig. \ref{lighttrends1}, \ref{lighttrends2}, \ref{lighttrends3}, \ref{lighttrends4} and \ref{lighttrends5}. The coefficient for the change in radiance ($\%$ per year) is listed in table \ref{table_trends} for each stop. Additional data is given for the whole city of Berlin and the stop at the very center (Alexanderplatz) The time series data is shown to allow the reader to form an opinion about the dynamics of night-time lights in the region we investigated. A comprehensive discussion is not the scope of this work.

\begin{table}[htbp]
  \centering
  \caption{Information about the measurement points.}
  \label{table_trends}
  \begin{tabular}{ccc}
	&& radiance change\\
stop Nr.& name & VIIRS/DNB\\
\hline 
- & Berlin & -1.2 $\%$/a\\
0 & Alexanderplatz &		+0.4 $\%$/a	\\
1 & Museumsinsel		&		-0.1 $\%$/a	\\
2 & Waldeckpark			&	-1.7 $\%$/a	\\
3 & Park am Gleisdreieck &			+0.1 $\%$/a	\\
4 & Hans Baluschek Park 	&		+0.8 $\%$/a	\\
5 & Gemeindepark Lankwitz	&	 +0.2 $\%$/a	\\
6 & Gut Osdorf			&		 -3.1 $\%$/a	\\
7 & Wietstock					&		-2.3 $\%$/a	\\
8 & Christinendorf &		+3.3 $\%$/a	\\
9 & Alexanderdorf &	+1.8 $\%$/a	\\
10 & Sperenberg	&		+3.3 $\%$/a	\\
11 & St{\"u}lpe 	&		+0.9 $\%$/a	\\
12 & Petkus	&	-1.0 $\%$/a	\\
  \end{tabular}
\end{table}

\end{document}